\documentclass[lettersize,journal]{IEEEtran}

\usepackage{amsmath,amsfonts}
\usepackage{algorithmic}
\usepackage{algorithm}
\usepackage{array}
\usepackage[caption=false,font=normalsize,labelfont=sf,textfont=sf]{subfig}
\usepackage{textcomp}
\usepackage{stfloats}
\usepackage{url}
\usepackage{verbatim}
\usepackage{graphicx}
\usepackage{cite}
\usepackage{xcolor}
\usepackage{colortbl}
\usepackage{booktabs}
\usepackage{multirow}
\usepackage{amsmath}
\usepackage{makecell}
\usepackage{colortbl}
\usepackage{orcidlink}
\usepackage[para,online,flushleft]{threeparttable}

\usepackage[nolist]{acronym}
\begin{acronym}
\acro{KWS}{Keyword Spotting}
\acro{IoT}{internet of things}
\acro{IC}{integrated circuit}
\acro{SoC}{system-on-chip}
\acro{RNN}{recurrent neural network}
\acro{CNN}{convolutional neural network}
\acro{DNN}{deep neural network}
\acro{IIR}{infinite impulse response}
\acro{BPF}{bandpass filter}
\acro{FEx}{feature extractor}
\acro{GSCD}{Google Speech Command Dataset}
\acro{AI}{artificial intelligence}
\acro{FFT}{Fast Fourier Transform}
\acro{SRAM}{static random-access memory}
\acro{DRAM}{dynamic random-access memory}
\acro{SPI}{serial peripheral interface}
\acro{FPGA}{field programmable gate array}
\acro{GRU}{gated recurrent unit}
\acro{FC}{fully-connected}
\acro{MAC}{multiply-accumulate}
\acro{SOS}{second-order sections}
\acro{MUX}{multiplexer}
\acro{BLK}{block}
\acro{ADC}{analog-to-digital converter}
\acro{MFCC}{Mel Frequency Cepstral Coefficients}
\acro{FoM}{figure of merit}
\end{acronym}

\usepackage{lipsum}
\usepackage{fancyhdr}
\fancypagestyle{sec}{\lhead{\tmpx}}

\fancypagestyle{arXiv}
{
   \fancyhf{}
   \chead{\textcolor{gray}{This paper has been accepted for publication in the IEEE Transactions on Circuits and Systems for Artificial Intelligence (TCASAI)}}
   \cfoot{ \fontsize{=9pt}{10pt}\selectfont  \textcolor{gray}{© 2025 IEEE. Personal use of this material is permitted. Permission from IEEE must be obtained for all other uses, in any current or future media, including reprinting/republishing this material for advertising or promotional purposes, creating new collective works, for resale or redistribution to servers or lists, or reuse of any copyrighted component of this work in other works}\\\fontsize{=10pt}{12pt}\selectfont\thepage}
   
}

\begin{document}

\title{DeltaKWS: A 65nm 36nJ/Decision Bio-inspired Temporal-Sparsity-Aware Digital Keyword Spotting IC with 0.6V Near-Threshold SRAM}
\author{Qinyu Chen*\orcidlink{0009-0005-9480-6164},~\IEEEmembership{Member,~IEEE}, 
Kwantae Kim*\orcidlink{0000-0001-8962-4554},~\IEEEmembership{Member,~IEEE}, 
Chang Gao*\orcidlink{0000-0002-3284-4078},~\IEEEmembership{Member,~IEEE}, 
Sheng Zhou\orcidlink{0009-0009-5408-5963}, 
Taekwang Jang\orcidlink{0000-0002-4651-0677},~\IEEEmembership{Senior Member,~IEEE}, 
Tobi Delbruck\orcidlink{0000-0001-5479-1141},~\IEEEmembership{Fellow,~IEEE}, 
Shih-Chii Liu\orcidlink{0000-0002-7557-045X},~\IEEEmembership{Fellow,~IEEE}
\thanks{{\textsuperscript{*}}Equal Contribution.}
\thanks{Qinyu Chen was with the Institute of Neuroinformatics, University of Zürich and ETH Zürich, Switzerland. She is now with the Leiden Institute of Advanced Computer Science (LIACS), Leiden University, The Netherlands.}
\thanks{Kwantae Kim was with the Institute of Neuroinformatics, University of Zürich and ETH Zürich, Switzerland, and also with the Department of Information Technology and Electrical Engineering (D-ITET), ETH Zürich, Switzerland. He is now with the Department of Electronics and Nanoengineering, School of Electrical Engineering, Aalto University, Espoo, Finland.}
\thanks{Chang Gao is with the Department of Microelectronics, Delft University of Technology, The Netherlands.}
\thanks{Taekwang Jang is with the Department of Information Technology and Electrical Engineering (D-ITET), ETH Zürich, Switzerland.}
\thanks{Sheng Zhou, Tobi Delbruck and Shih-Chii Liu are with the Institute of Neuroinformatics, University of Zürich and ETH Zürich, Switzerland.}
\thanks{
Corresponding authors: Qinyu Chen (q.chen@liacs.leidenuniv.nl) and Shih-Chii Liu (shih@ini.uzh.ch).
}
}

\markboth{}%
{Shell \MakeLowercase{\textit{et al.}}: A Sample Article Using IEEEtran.cls for IEEE Journals}


\maketitle

\begin{abstract}
This paper introduces DeltaKWS, to the best of our knowledge, the first $\Delta$RNN-enabled fine-grained temporal sparsity-aware \ac{KWS} \ac{IC} for voice-controlled devices. The 65\,nm prototype chip features a number of techniques to enhance performance, area, and power efficiencies, specifically:
1) a bio-inspired delta-gated recurrent neural network ($\Delta$RNN) classifier leveraging temporal similarities between neighboring feature vectors extracted from input frames and network hidden states, eliminating unnecessary operations and memory accesses; 
2) an \ac{IIR} \ac{BPF}-based \ac{FEx} that leverages mixed-precision quantization, low-cost computing structure and channel selection;
3) a 24\,kB 0.6\,V near-$V_\text{TH}$ weight \ac{SRAM} that achieves 6.6× lower read power than the foundry-provided \ac{SRAM}.
From chip measurement results, we show that the DeltaKWS achieves an 11/12-class \ac{GSCD} accuracy of 90.5\%/89.5\% respectively and energy consumption of 36\,nJ/decision in 65\,nm CMOS process. At 87\% temporal sparsity, computing latency and energy/inference are reduced by 2.4×/3.4×, respectively.
The IIR BPF-based FEx, $\Delta$RNN accelerator, and 24\,kB near-$V_\text{TH}$ \ac{SRAM} blocks occupy 0.084\,mm\textsuperscript{2}, 0.319\,mm\textsuperscript{2}, and 0.381\,mm\textsuperscript{2} respectively (0.78\,mm\textsuperscript{2} in total). 
\end{abstract}




\section{Introduction}
\thispagestyle{arXiv}
\IEEEPARstart{K}{eyword} Spotting (KWS) is an essential always-on function for voice-activated devices. 
\acresetall 
The KWS wakes up the downstream building blocks of the \ac{IoT} devices by recognizing specific keywords, facilitating ultra-low-power consumption and always-on operation. This area has attracted significant research attention for \ac{IC} in design community \cite{Kim2022JSSC, Shan2020ISSCC, Dbouk2021JSSC, Giraldo2020JSSC, Kosuge2023VLSI, tan2023, cai2024, Seol2023} where the challenge lies in minimizing power consumption while keeping high \ac{KWS} accuracy.

Advancements in \ac{AI} techniques have led to extensive applications of neural network models in \ac{KWS} classifiers. Models such as \acp{CNN}, \acp{RNN}, and attention mechanism networks utilize their powerful data mining capabilities to learn and identify keyword-relevant characteristics, resulting in superior accuracy. However, the complexity of these networks leads to redundant computations and increased hardware costs at the edge.

Reported methods to increase the energy efficiency of \ac{KWS} \acp{IC} include exploiting the inherent weight and activation sparsity of the subsequent neural network to reduce redundant computations. 
\ac{KWS} chip designs are commonly tested on the \ac{GSCD}. 
A recurrent attention in-memory \ac{KWS} \ac{IC} exploited activation sparsity but only achieved up to 70\% activation sparsity on a 7-class \ac{GSCD} subset \cite{Dbouk2021JSSC}. A switch-capacitor array-based \ac{KWS} chip used a \ac{CNN} model as a classifier, and exploited bit-level sparsity in weights and achieved 90.9\% accuracy on an 11-class \ac{GSCD} subset \cite{tan2023}. 
A \ac{KWS} \ac{SoC} running skip-\ac{RNN} exploited 76\% coarse-grained temporal sparsity by skipping audio frames but reported only on a 7-class \ac{GSCD} subset \cite{Seol2023}. 
None have exploited temporal sparsity in the change of neural network activations as an additional method of increasing the energy efficiency of \ac{KWS} \acp{IC}.

Other \acp{IC} focused on reducing the power of the audio \ac{FEx}, commonly employing serial \ac{FFT}-based or time-domain band-pass filtered features, and using a single deep neural network for both feature extraction and classification. 
A serial \ac{FFT}-based \ac{KWS} chip achieved 510\,nW but at the cost of a 64\,ms latency in a 2-class \ac{GSCD} subset \cite{Shan2020ISSCC}. A 23$\mu$W \ac{KWS} chip demonstrated an analog ring-oscillator-based time-domain \ac{FEx} but only achieved around 86\% accuracy on the full 12-class \ac{GSCD} \cite{Kim2022JSSC}. 
A single-chip fully synthesizable Wired-Logic \ac{DNN} processor \cite{Kosuge2023VLSI} used a single 16-layer \ac{DNN} model for feature extraction and classification; however, it only achieved 88\% accuracy for 10 keywords subset. 
In addition, 40\%-to-60\% \cite{tan2023, Kim2022JSSC, Shan2020ISSCC} of the total \ac{IC} power was dominated by weight memory (e.g., \ac{SRAM}), which limits the scaling to larger networks required for real-world applications.


We present DeltaKWS, which shows improvements in area and energy efficiency while achieving $>$ 90\% accuracy through the use of the following three key blocks:

\begin{enumerate}
    \item A bio-inspired $\Delta$\ac{RNN} accelerator leveraging temporal similarities between neighboring feature vectors extracted from input frames and network hidden states, eliminating unnecessary operations and memory accesses; achieving 2.4$\times$ latency reduction and 3.4$\times$ energy per decision reduction.
    \item A serial \ac{IIR}-\ac{BPF} \ac{FEx} featuring mixed-precision quantization, low-cost computing structure, and channel selection, achieving 5.7$\times$ power reduction and 4.7$\times$ area reduction.
    \item A 24\,kB 0.6\,V near-threshold-operating full-custom \ac{SRAM} achieving 6.6$\times$ lower power consumption.
    \end{enumerate}

The proposed \ac{KWS} chip prototyped in 65\,nm CMOS process incorporates an \ac{IIR} \ac{BPF}-based \ac{FEx}, $\Delta$\ac{RNN} accelerator, and 24\,kB near-threshold \ac{SRAM}, occupying 0.78\,mm\textsuperscript{2}, with 36\,nJ/Decision energy efficiency.

This article is organized as follows. 
Section~\ref{sec:SKWS} details the bio-inspired temporal sparsity-aware KWS system. 
Section~\ref{sec:results} reports the system implementation of the always-on KWS task
and measurement results. 
Section~\ref{sec:conclusion} concludes this article.

\begin{figure}[t]
    \centering
    \includegraphics[width=0.5\textwidth]{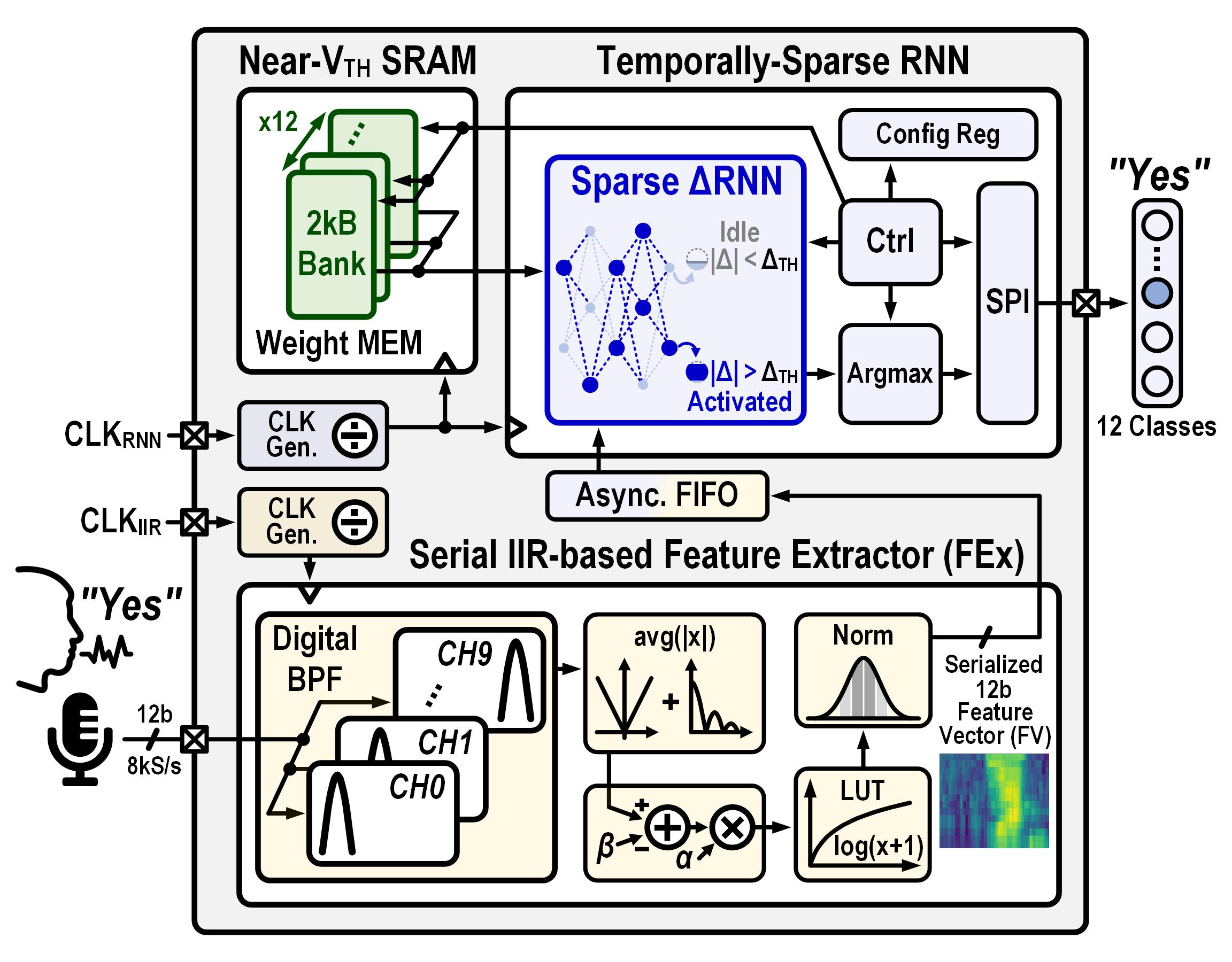}
    \caption{Overall architecture of proposed temporally-sparse $\Delta$\ac{RNN} \ac{KWS} \ac{IC} with \ac{IIR} \ac{BPF}-based time-domain \ac{FEx} and near-$V_\text{TH}$ weight \ac{SRAM}.}
    \label{fig:1}
    \vspace{-1mm}
\end{figure}

\section{Bio-inspired Temporal-Sparsity-Aware \ac{KWS} System}
\label{sec:SKWS}

The details of the DeltaKWS architecture are described first in Section~\ref{sec:arch}. Section~\ref{sec:DRNNarch} describes the blocks of the $\Delta$RNN accelerator. 
Section~\ref{sec:IIR} covers the IIR BPF-based FEx, and Section~\ref{sec:SRAM} provides information about the near-$V_\text{TH}$ \ac{SRAM}.

\subsection{Architectural Overview}
\label{sec:arch}
Fig.~\ref{fig:1} shows the chip architecture with a $\Delta$\ac{RNN} accelerator, a 24\,kB near-$V_\text{TH}$ \ac{SRAM} for weight storage, and a compact serial \ac{IIR} \ac{BPF}-based time-domain \ac{FEx}.
The input of \ac{FEx} is 12b streamed with a \ac{SPI} interface.  
An asynchronous FIFO connects \ac{FEx} 12b outputs to the $\Delta$\ac{RNN} accelerator across various clock domains. 
Two clock dividers driven by the master clock generated by a \ac{FPGA} are used to reduce the fast clocks ($\text{CLK}_\text{RNN}$ and $\text{CLK}_\text{IIR}$) to slower speeds. This is necessary because while \ac{SPI} transfers 1b per clock cycle and requires a faster clock for interfacing, the processing within the chip operates more efficiently at a slower clock speed.
The 24\,kB near-$V_\text{TH}$ \ac{SRAM} stores the entire $\Delta$\ac{RNN} model weights. On-chip storing avoids constant off-chip weight access, saving considerable energy (1\,$\mu$J to read 6.2\,kB data from \ac{DRAM} \cite{Horowitz2014}).

\subsection{Temporally Sparse $\Delta$\ac{RNN}} 
\label{sec:DRNNarch}
\begin{figure}[t]
    \centering
    \includegraphics[width=0.36\textwidth]{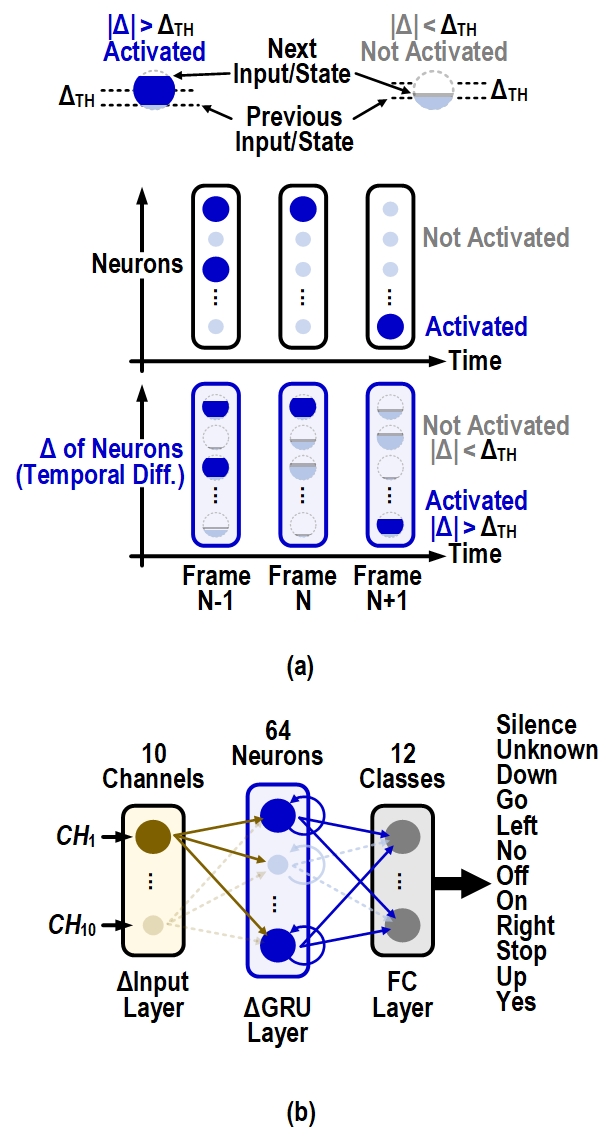}
    \caption{Concept of $\Delta$ network: (a) $\Delta$ neurons along time axis. The bottom row illustrates the process by which neuron states are determined across frames in a GRU network, showing the temporal differences that drive these state changes. The middle row shows the resulting neuron states across frames in the GRU network, indicating which neurons are activated (dark blue) or inactivated (light blue) based on these temporal differences. The upper row provides a detailed view, indicating that a neuron is only activated when its temporal difference exceeds the defined threshold. (b) The $\Delta$\ac{GRU} structure, with 10 input channels, processes data through a $\Delta$\ac{RNN} layer containing 64 neurons, followed by a fully connected layer that classifies inputs into 12 command categories: ‘Silence,’ ‘Unknown,’ ‘Down,’ ‘Go,’ ‘Left,’ ‘No,’ ‘Off,’ ‘On,’ ‘Right,’ ‘Stop,’ ‘Up,’ and ‘Yes.’ In this work, we also evaluate 11-class accuracy \cite{tan2023}, excluding the ‘Unknown’ category.}
    \label{fig:drnn_0}
    \vspace{-1mm}
\end{figure}

\begin{figure}[t]
    \centering
    \includegraphics[width=0.45\textwidth]{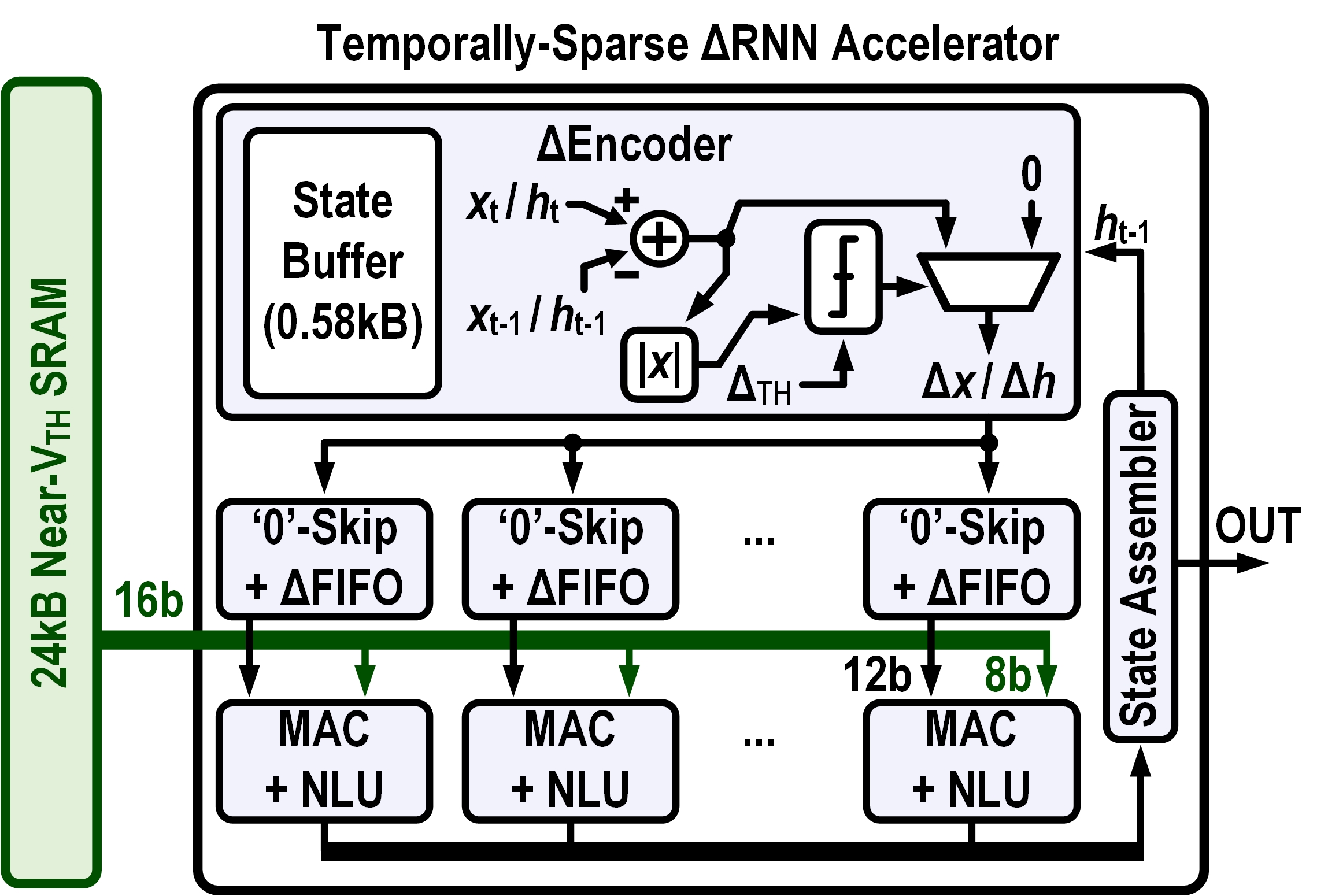}
    \caption{The architecture of the $\Delta$\ac{RNN} accelerator. The 24\,kB weight memory provides 16\,bit words, storing two 8\,bit $\Delta$\ac{RNN} weights.}
    \label{fig:drnn_1}
    \vspace{-1mm}
\end{figure}



Neurons in the cerebral cortex exhibit a spiking threshold; their activation is not only discrete over time but also capped at a finite firing rate. Inspired by these neuromorphic characteristics, researchers have explored the concept of inducing temporal sparsity in artificial recurrent neural networks. This approach has been termed the $\Delta$\ac{RNN} model~\cite{Neil2017ICML,Gao2018FPGA,Gao2020JETCAS,Liu2022IEDM,Ottati2023JETCAS}. 
Fig.~\ref{fig:drnn_0} illustrates the concept of the $\Delta$\ac{RNN} and the architecture for the \ac{KWS} network. 
The $\Delta$\ac{RNN} employs fine-grained temporal sparsity by allowing neurons to update their targets only when their delta change in activation exceeds a defined threshold, thus skipping unnecessary computations and memory accesses. This mechanism can significantly boost energy efficiency. The \ac{KWS} network architecture comprises a $\Delta$Input layer, a $\Delta$Gated Recurrent Unit ($\Delta$GRU) for the hidden layer, and a final \ac{FC} layer that classifies 12 \ac{GSCD} classes. 

The $\Delta$\ac{RNN} hardware accelerator is illustrated in Fig.~\ref{fig:drnn_1}. It features a $\Delta$Encoder for calculating temporal differences of both the input activations ($\Delta x=x_t-x_{t-1}$) and hidden states ($\Delta h=h_t-h_{t-1}$), focusing on critical neuron updates that exceed the set delta threshold ($\Delta_{\mathrm{TH}}$) to optimize the trade-off between accuracy and energy consumption. The accelerator broadcasts each non-zero delta state to all delta FIFOs ($\Delta$FIFOs) and utilizes eight \ac{MAC} units for matrix-vector multiplication calculations. A State Assembler ensures proper update and coordination of $\Delta$\ac{RNN} states. 



\subsection{\ac{IIR} \ac{BPF}-based \ac{FEx}}
\label{sec:IIR}
\begin{figure}[t]
    \centering
    \includegraphics[width=0.48\textwidth]{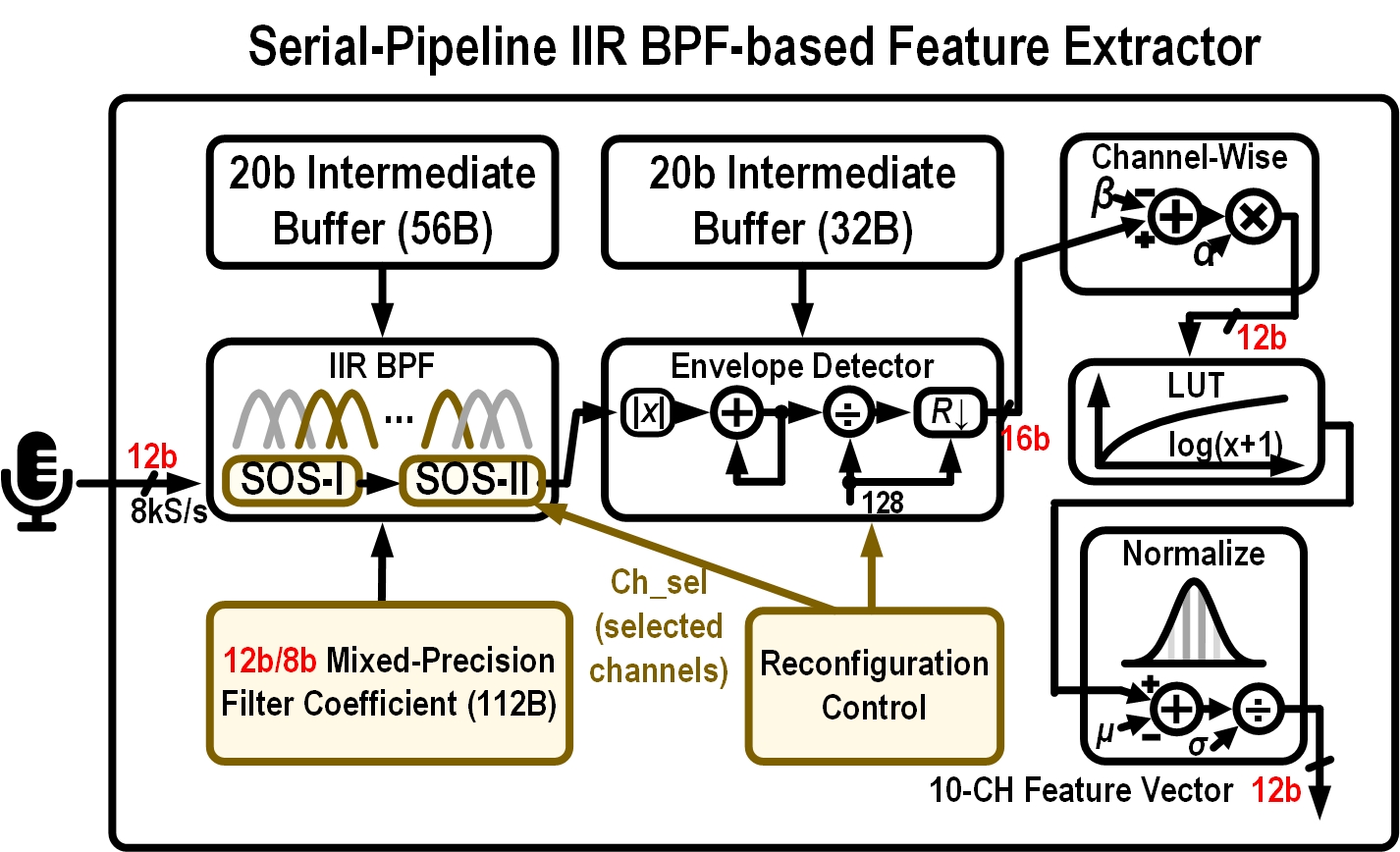}
    \caption{The architecture of the serial time-domain \ac{IIR} \ac{BPF}-based \ac{FEx}\textcolor{red}{.}}
    \label{fig:iir_0}
    \vspace{-1mm}
\end{figure}

\begin{figure}[t]
    \centering
    \includegraphics[width=0.45\textwidth]{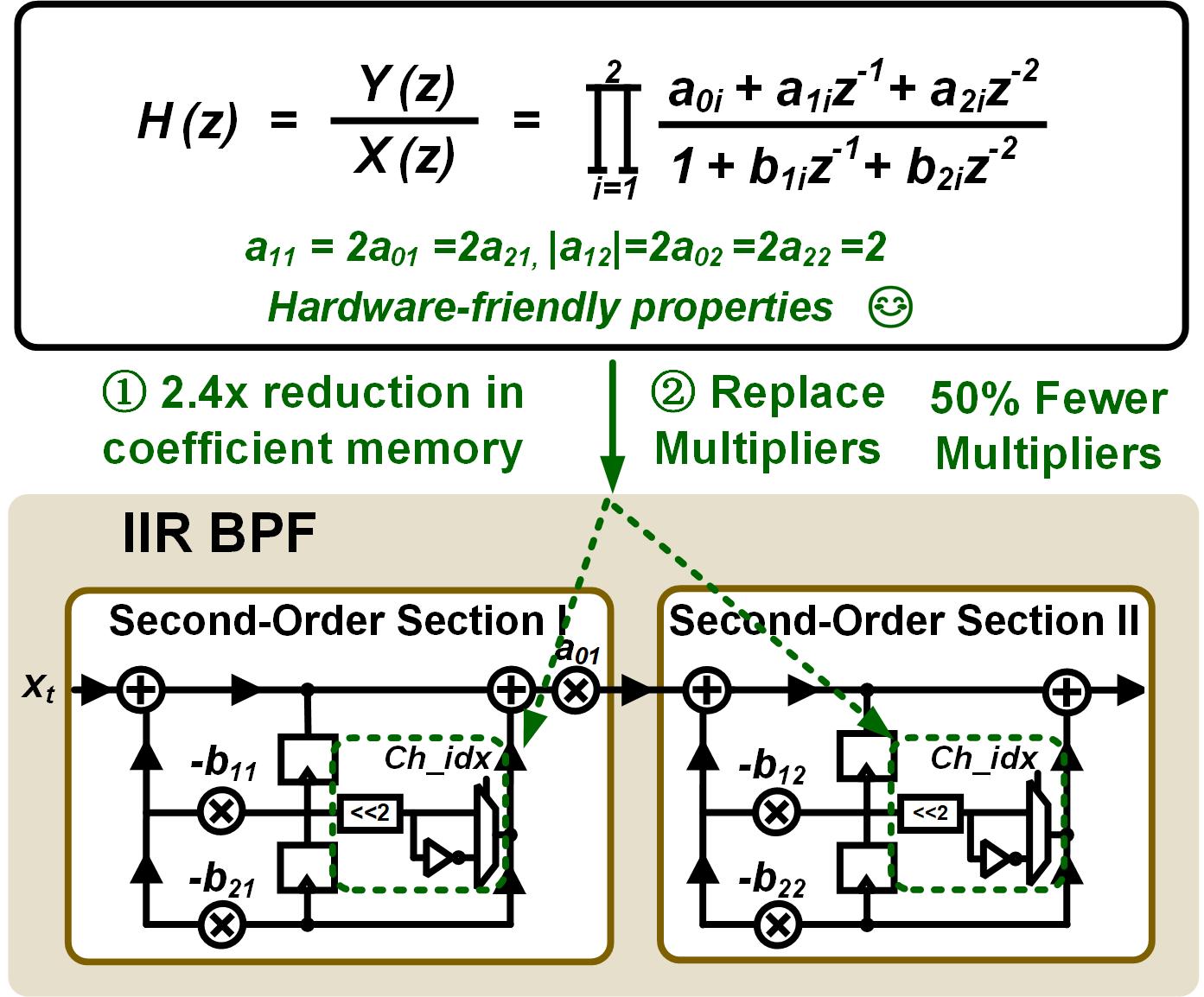}
    \caption{Low arithmetic complexity structure of the 4\textsuperscript{th} order \ac{IIR} \ac{BPF}.}
    \label{fig:iir_1}
    \vspace{-1mm}
\end{figure}

\begin{figure}[t]
    \centering
    \includegraphics[width=0.5\textwidth]{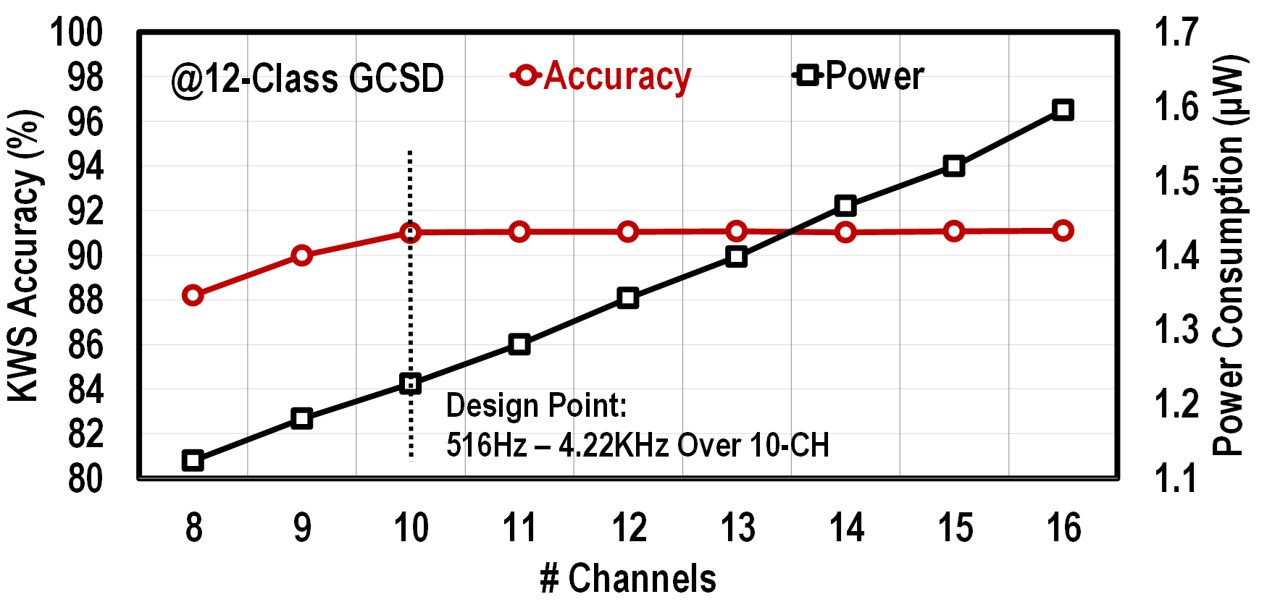}
    \caption{Simulated power versus 12-class \ac{KWS} accuracy over the different number of channels.}
    \label{fig:iir_2}
    \vspace{-1mm}
\end{figure}

\begin{figure}[ht]
    \centering
    \includegraphics[width=0.35\textwidth]{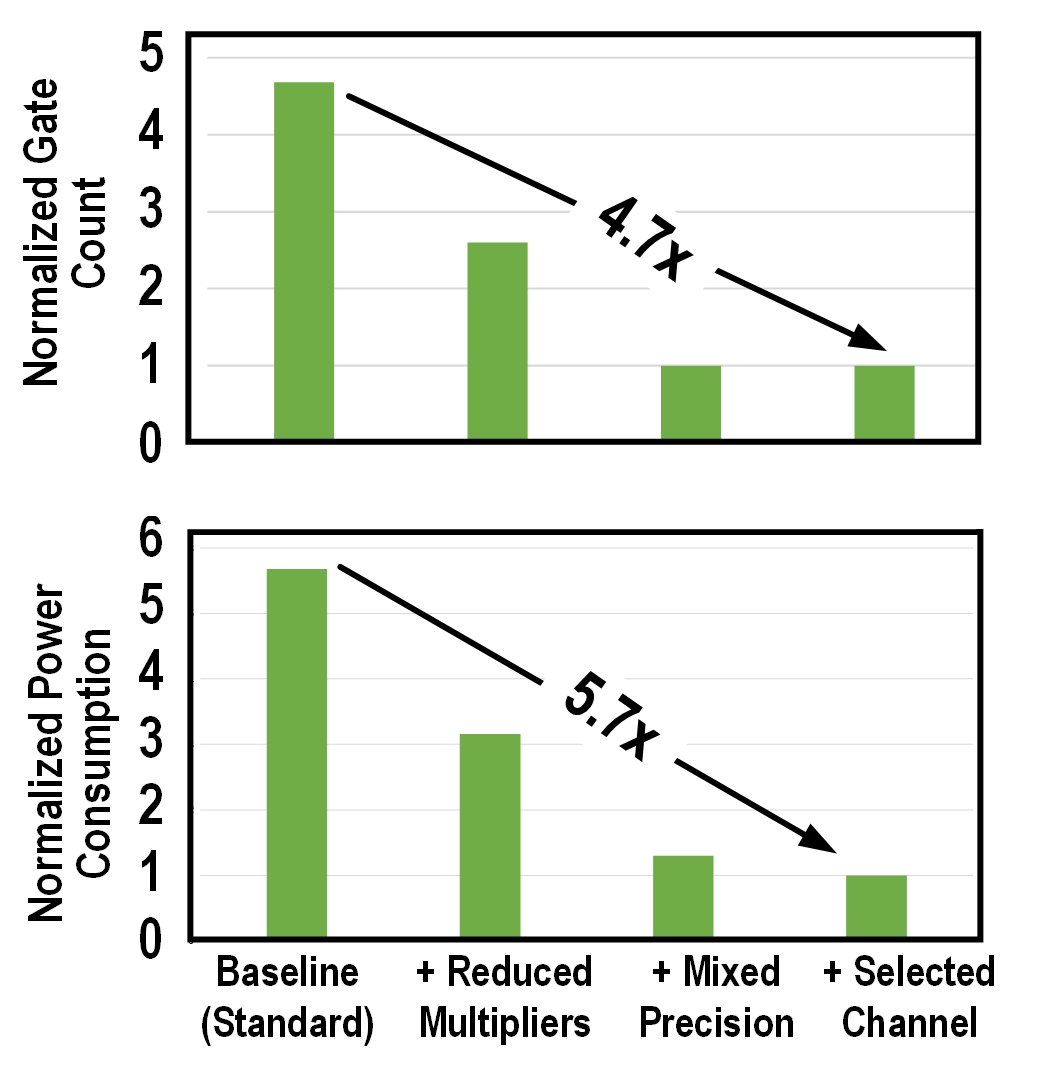}
    \caption{The area (gate count) and power consumption optimization over each step.}
    \label{fig:iir_3}
    \vspace{-1mm}
\end{figure}


Fig.~\ref{fig:iir_0} describes the serial \ac{IIR} \ac{BPF}-based FEx~\cite{Chen2023ISCAS} architecture generating 12b feature vectors to the $\Delta$\ac{RNN} accelerator.
The \ac{FEx} includes 1) a 4\textsuperscript{th}-order \ac{IIR} \ac{BPF}, implemented as a cascade of two \ac{SOS}, 2) a post-processing unit, which implements an envelope detector, channel-wise offset and scale adjustments, log compression, and normalization, 3) register files storing intermediate results and filter coefficients, and 4) a reconfiguration control module to select the channels for processing.  
The input \ac{GSCD} speech samples are quantized to 12b and sub-sampled at 8\,kHz before they are presented to the \ac{FEx}. 

\subsubsection{Reduced Arithmetic Complexity in IIR BPF Design}
Fig.~\ref{fig:iir_1} shows the proposed low-arithmetic-complexity 4\textsuperscript{th} order \ac{IIR} \ac{BPF} computation structure. 
According to the basic architecture of a 4\textsuperscript{th} order \ac{IIR} \ac{BPF}, 10 multipliers and 8 adders are required. 
If we look closer at coefficients $a$ and $b$, we see hardware-friendly properties such as symmetries and constant value representations. 
By exploiting the hardware-friendly properties, half of the multipliers can be replaced with bit shifts.

\subsubsection{Reconfigurable Channel Selection} 
The \ac{FEx} is designed to be reconfigurable, supporting configurations from 1 to 16 channels. 
The simulated power versus \ac{KWS} accuracy over different number of channels is shown in Fig.~\ref{fig:iir_2}. The simulation results show that the 12-class \ac{KWS} accuracy with \ac{GSCD} is maintained even when the number of \ac{IIR} filter channels decreases to 10 channels covering a frequency range from 516\,Hz to 4.22\,kHz. 
In the hardware design, the reconfiguration module adjusts the number of channels to be computed. In this case, selecting 10 channels instead of 16 reduces the power consumption of the \ac{FEx} by 30\%. In addition, the reconfigurable design enables our hardware to support up to 16 channels, providing flexibility for other applications that might benefit from a larger frequency range.

\subsubsection{Mixed-Precision Selection of Filter Coefficients}

Due to potential numerical stability issues, a fixed-point analysis is required during the \ac{IIR} filter implementation. The selection of coefficient precision is essential for power and area reduction. 
Previous works adopted the unified bit precision for coefficients. Thanks to the error resilience of neural networks, we can select an aggressive bit-width selection, in other words, the network accuracy is used as the measurement standard.
Through a grid search, we find that the dynamic ranges for coefficients $a$ and $b$ across the filters of \ac{FEx} are different, and the impact of bit precision on \ac{KWS} accuracy also varies. The independent bit precision selection allows for further reduction of power and hardware resources. 
The integer bits for $a$ and $b$ are first determined separately using their maximum values. 
The fraction bits are then reduced from the baseline (16-bit) and the network accuracy is again quantified.   
From this optimization step, we find that 12b/8b \textit{(b/a)} mixed precision is sufficient for the desired accuracy.

Fig.~\ref{fig:iir_3} depicts the power consumption and area (gate count) optimization over each optimization step.
The filter coefficients are realized with 12b/8b mixed-precision, obtaining 2.4$\times$ power and 2.6$\times$ area reduction. Furthermore, the hardware-friendly properties (i.e., equivalence) of $a$ and $b$ coefficients of biquad filters are exploited so that half of the multipliers can be replaced with bit shifts, leading to 1.8$\times$ power and 1.8$\times$ area reduction. Putting them all together, the power and area of the proposed digital \ac{FEx} are reduced by 5.7$\times$ and 4.7$\times$, respectively.

Unlike most prior arts~\cite{Shan2020ISSCC,Shan2022aadkws, zheng2019kws} relying on 16b high-resolution \acp{ADC}, the proposed \ac{IIR} digital \ac{FEx} features a 12b-quantized audio signal, facilitating the usage of low-power/low-area 12b \acp{ADC}.

\subsection{Near-$V_{TH}$ Weight \ac{SRAM}}
\label{sec:SRAM}

\begin{figure*}[t]
    \centering
    \includegraphics[width=\textwidth]{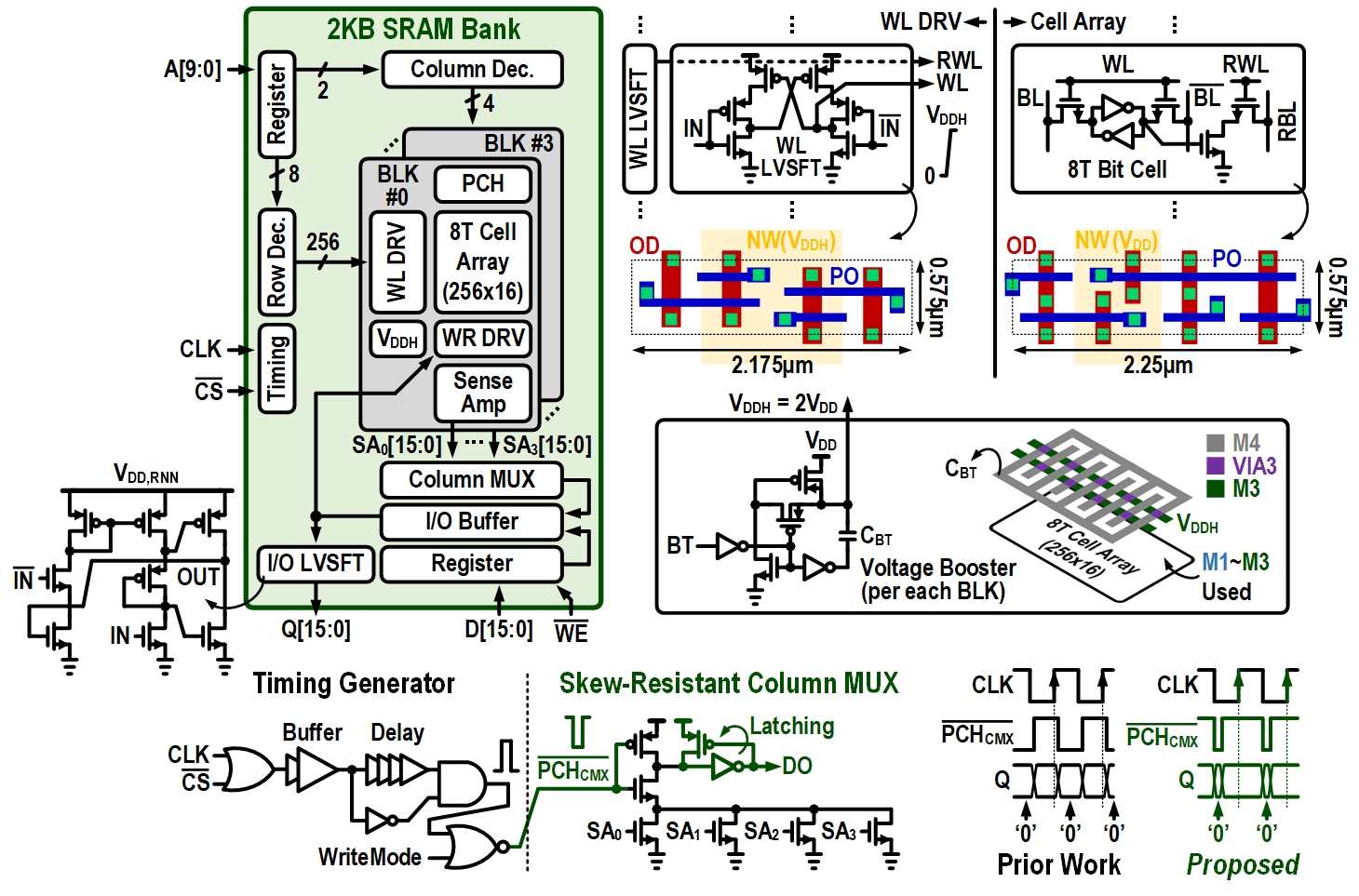}
    \caption{0.6V near-$V_\text{TH}$ full-custom \ac{SRAM} with pitch-matched WL level shifter, I/O level shifter, integrated voltage booster, timing generator, and skew-resistant column \ac{MUX}.}
    \label{fig:4}
    \vspace{-1mm}
\end{figure*}

Fig.~\ref{fig:4} illustrates a 24\,kB weight \ac{SRAM} operating at the 0.6\,V near-$V_\text{TH}$ voltage, which is divided into 12 banks of 2\,kB units each. The \ac{SRAM} comprises 4 \acp{BLK} of memory, a 10b address (A) register, a 16b input (D) register, a 0.6\,V-to-0.65\,V output (Q) level shifter for $\Delta$\ac{RNN} interfacing, and a skew-resistant column \ac{MUX}. Note that the 0.6\,V/0.65\,V dual-supply approach was chosen only due to the low-to-high nature of I/O level shifter \cite{Lotfi2018lvsft}. While not addressed in the design phase of this chip, a single supply operation is possible if the Q output bypasses I/O level shifters or by adopting a bidirectional level shifter \cite{Luo2014lvsft}. The proposed \ac{SRAM} deploys pitch-matched 6T level shifters that align with the 8T bitcell’s vertical pitch to convert word lines (WL/RWL) from 0.6\,V to 1.2\,V leading to enhanced access time and static noise margin.

Unlike prior art \cite{Chang2008ISSCC} relying on additional off-chip above-$V_\text{DD}$ WL booster, the proposed \ac{SRAM} features an on-chip-integrated voltage booster ($V_\text{DDH}$)~\cite{Verma2007ISSCC}. 
Our combination approach of pitch-matched level shifter and voltage booster is distinct from \cite{Verma2007ISSCC}, which utilized the booster circuit only for driving the footer of the read-buffer within the 8T bitcell, rather than word lines.

The \ac{SRAM} operates at reduced leakage power by employing high-$V_\text{TH}$ devices for bitcells. The column \ac{MUX} uses dynamic NOR logic~\cite{Gupta2019TCAS-I} and the proposed clock-skew-robust pre-charging scheme (PCHCMX), ensuring output data (Q) refreshes at the falling clock edge and facilitating easier integration of full-custom design with synthesized logic. The implemented full-custom \ac{SRAM} achieves a 6.6$\times$ lower read power of 0.93\,$\mu$W (only 18\% of the total power of the chip) and a 2$\times$ larger area than the push-rule foundry \ac{SRAM} \cite{Kim2022ISSCC}.

\section{Measurement Results}
\label{sec:results}
\begin{figure}[t]
    \centering
    \includegraphics[width=0.4\textwidth]{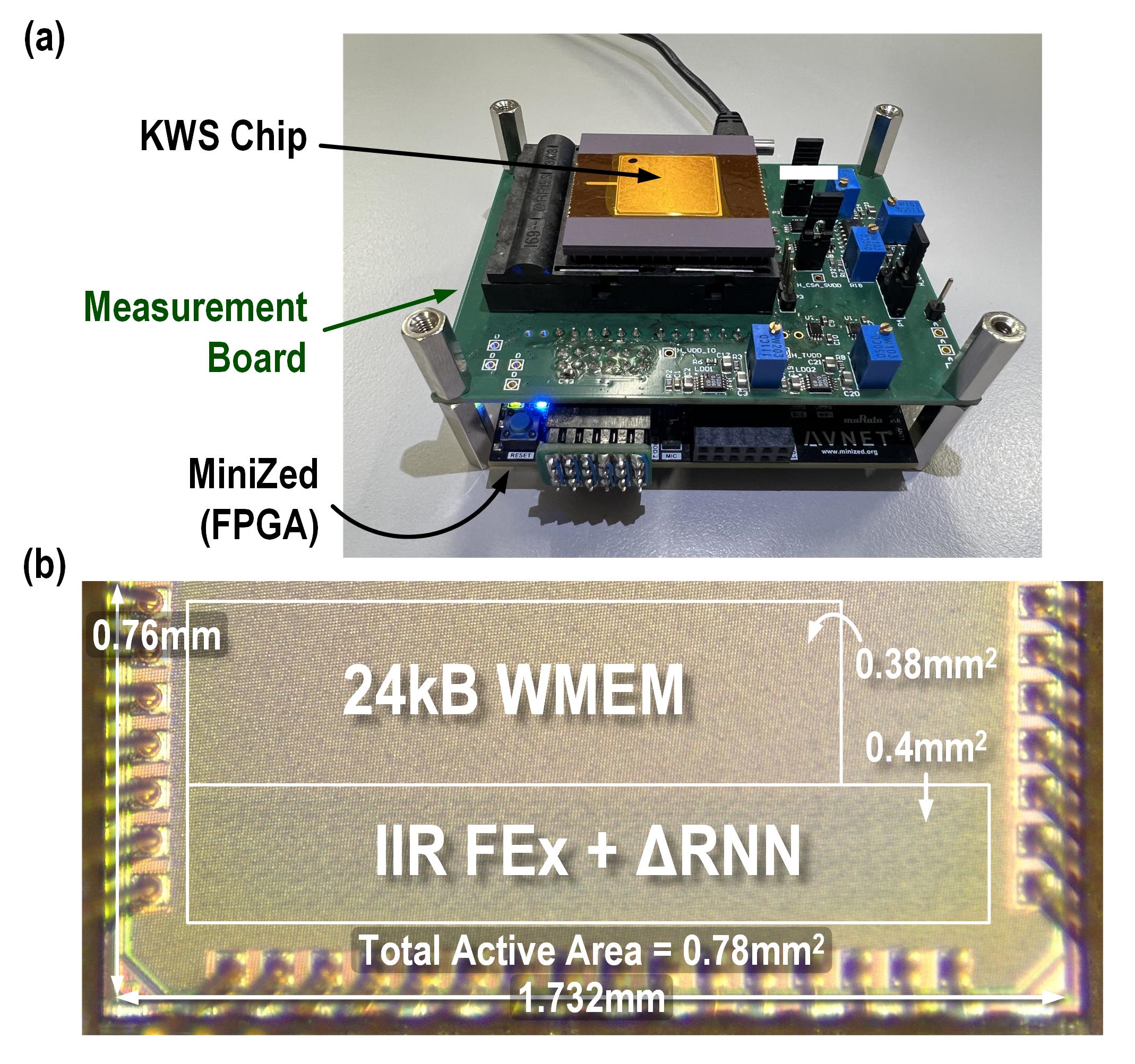}
    \caption{(a) Measurement setup of the KWS chip. (b) Chip micrograph.}
    \label{fig:chip}
    \vspace{-1mm}
\end{figure}

\begin{figure}[t]
    \centering\includegraphics[width=0.5\textwidth]{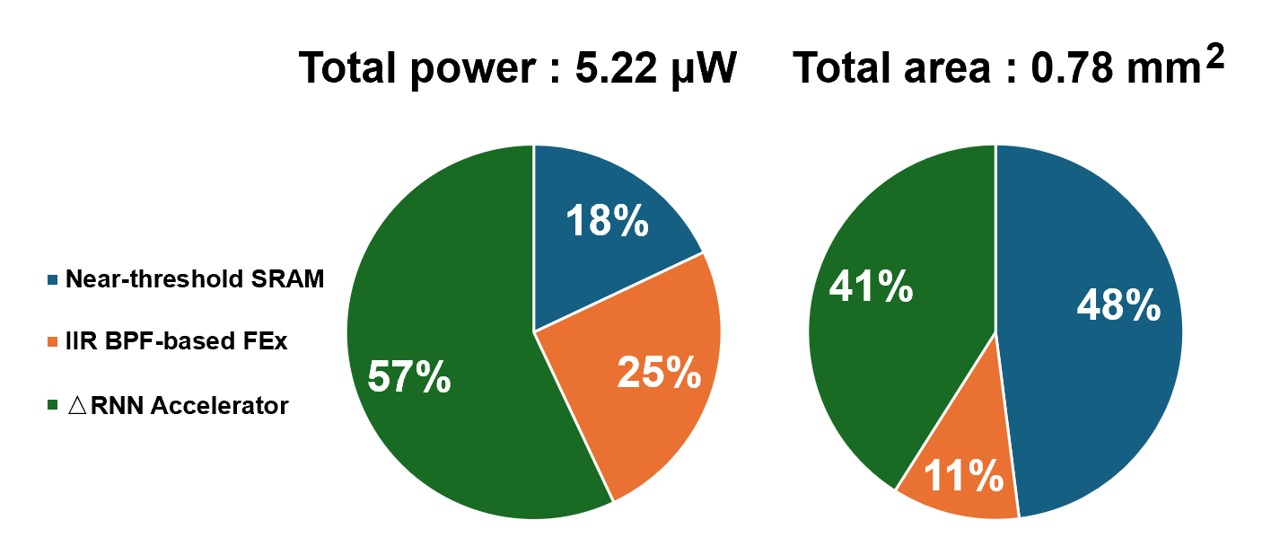}
    \caption{Measured power and area breakdown of the KWS chip.}
    \label{fig:power_breakdown}
    \vspace{-1mm}
\end{figure}

\begin{figure}[t]
    \centering
    \includegraphics[width=0.4\textwidth]{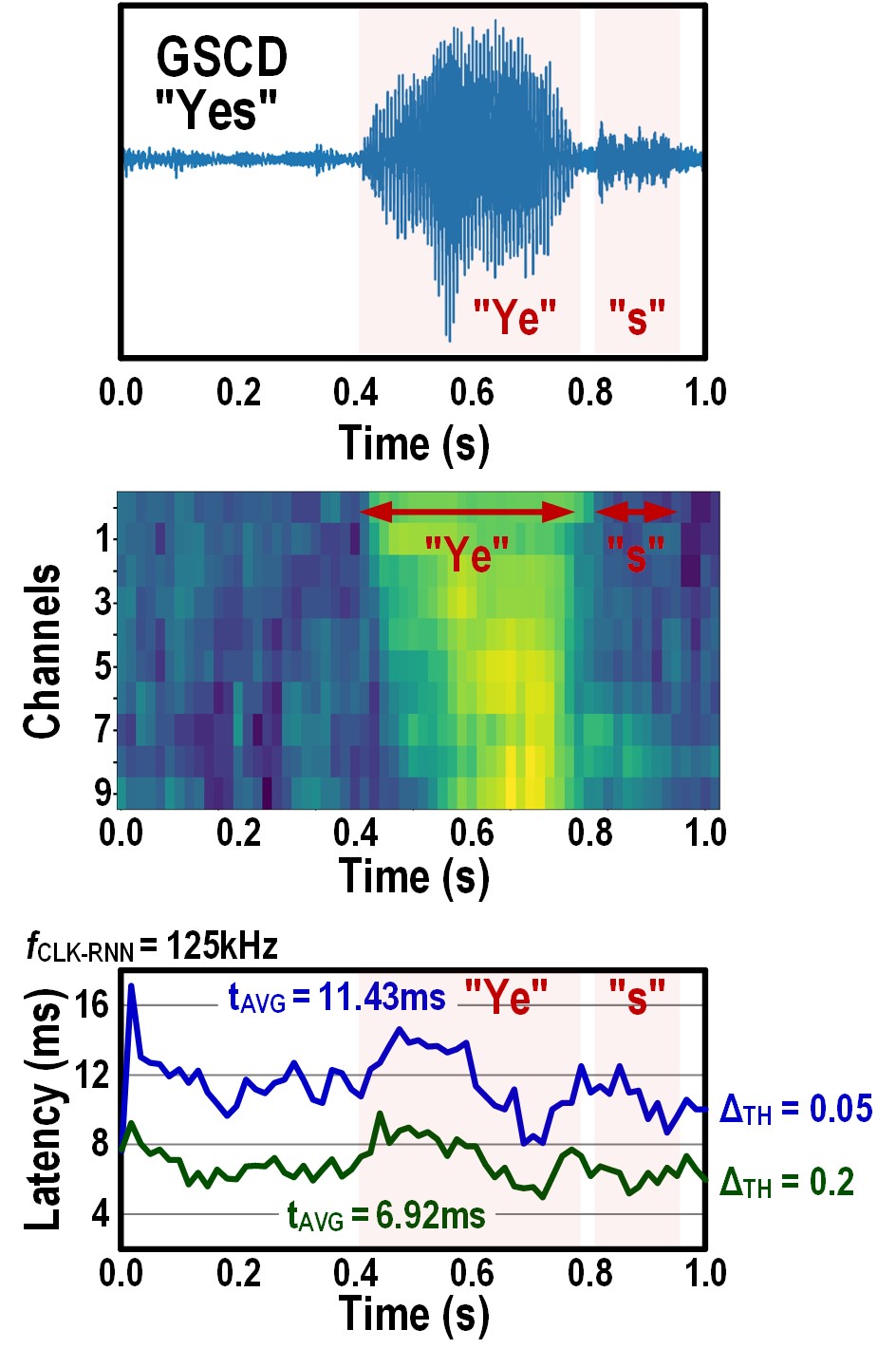}
    \caption{(Top) audio waveform of “Yes” sample;  (middle) measured IIR features and (bottom) $\Delta$\ac{RNN} latency for two $\Delta_{TH}$ values.}
    \label{fig:measure_0}
    \vspace{-1mm}
\end{figure}


\begin{figure}[t]
    \centering
    \includegraphics[width=0.4\textwidth]{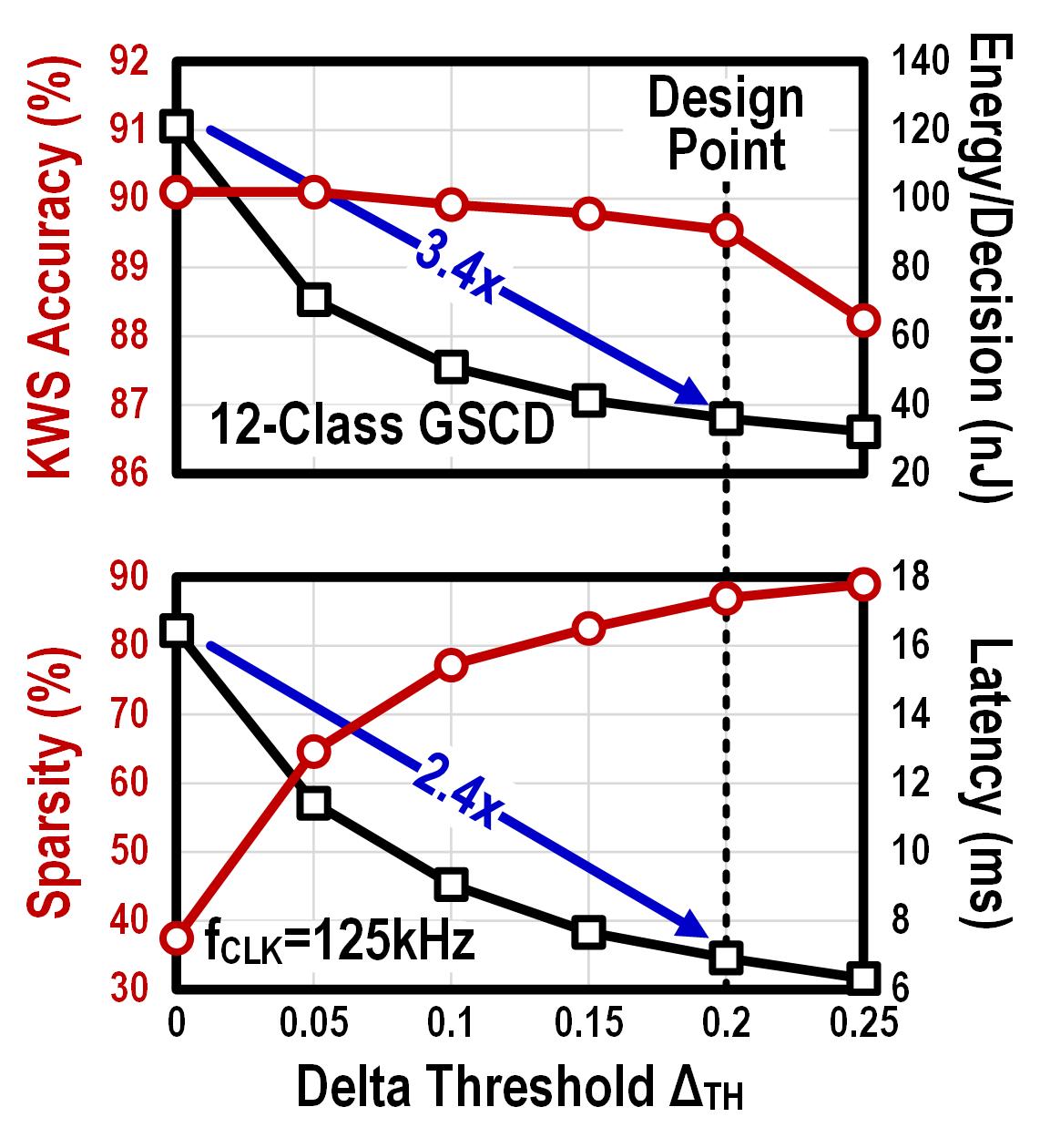}
    \caption{Measured 12-class \ac{GSCD} \ac{KWS} accuracy, energy per decision, average temporal sparsity and computing latency at different delta thresholds $\Delta_\text{TH}$. 125kHz clock is used. $\Delta_\text{TH}=0.2$ is chosen as the design point, with 3.4x lower energy and 2.4x shorter latency at 87\% sparsity.}
    \label{fig:measure_1}
    \vspace{-1mm}
\end{figure}

\begin{figure}[t]
    \centering
    \includegraphics[width=0.35\textwidth]{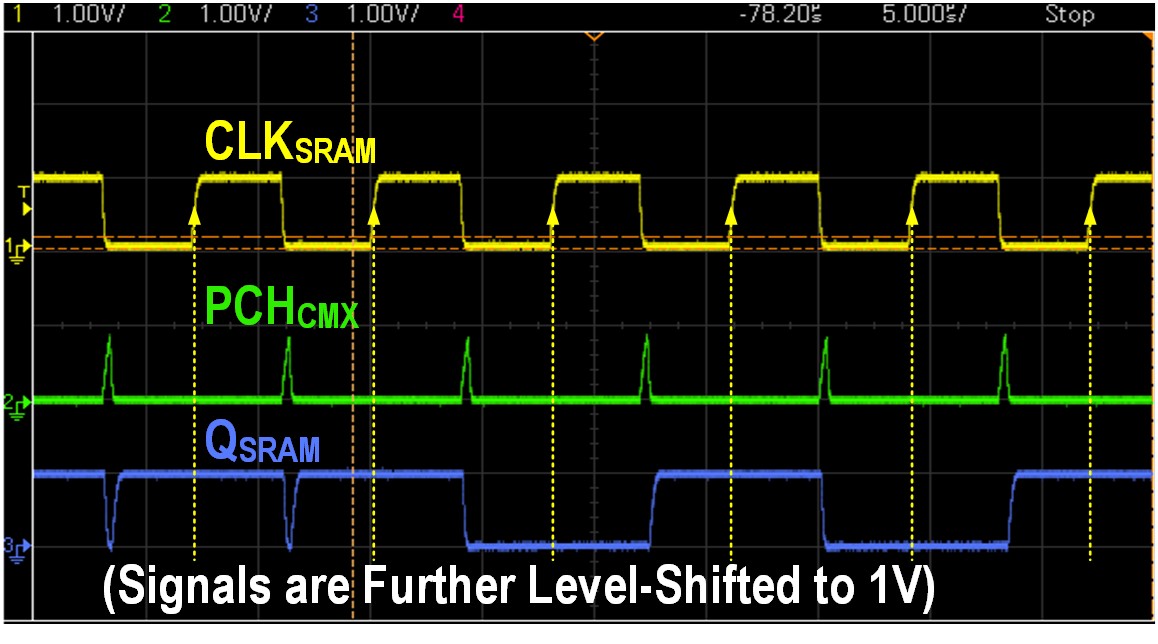}
    \caption{Skew-resistant column MUX measurement from SRAM.}
    \label{fig:measure_3}
    \vspace{-1mm}
\end{figure}

The die photo and measurement setup of the DeltaKWS chip fabricated in a 65\,nm CMOS technology are shown in Fig.~\ref{fig:chip}.
Fig.~\ref{fig:power_breakdown} summarizes the power and area breakdown at 125\,kHz operating frequency. 
The \ac{KWS} chip consumes 5.22\,$\mu$W. The \ac{IIR} \ac{BPF}-based \ac{FEx}, $\Delta$\ac{RNN} accelerator, and near-$V_\text{TH}$ \ac{SRAM} consume 11\%, 41\%, and 48\% of the total power, respectively. The \ac{IIR} \ac{BPF}-based \ac{FEx}, $\Delta$\ac{RNN} accelerator, and near-$V_\text{TH}$ \ac{SRAM} consume 25\%, 57\%, and 18\% of the total power, respectively.
The chip core area is 0.78\,mm\textsuperscript{2}. 
A MiniZed development board with a Xilinx Zynq 7007S \ac{SoC} \ac{FPGA} is used as a host to initialize and feed inputs to the \ac{KWS} \ac{IC}.
To validate the \ac{KWS} chip's performance, we downsample the test \ac{GSCD} audio streams to 8\,kHz. The downsampling induces a negligible influence on the \ac{GSCD} accuracy.

\subsection{KWS Accuracy, Energy, and Latency}
Fig.~\ref{fig:measure_0} shows the audio waveform of a 1-second \ac{GSCD} “Yes” keyword and the measurement results on \ac{IIR} features and the $\Delta$\ac{RNN} latency.
The \ac{FEx} produces 12-bit 10-channel features from \ac{IIR} \acp{BPF} with Mel-scale center frequencies from 516\,Hz to 4.22\,kHz, that are fed to the $\Delta$\ac{RNN} accelerator through an asynchronous FIFO.
Relatively silent frames in the “Yes” sample lead to around 40\% latency reduction compared to active frames.

The temporal sparsity in the $\Delta$\ac{RNN} can by adjusted through $\Delta_\text{TH}$. The level of sparsity also impacts the task accuracy; and the network energy and latency per input sample. 
Fig.~\ref{fig:measure_1} shows the measured 12-class \ac{GSCD} \ac{KWS} accuracy, energy per decision, average temporal sparsity and computing latency of the KWS chip at various $\Delta_\text{TH}$.
In comparison to $\Delta_\text{TH}$ = 0, $\Delta_\text{TH}$ can be increased to 0.2 to achieve 87\% temporal sparsity while maintaining 89.5\% accuracy on the 12-class \ac{GSCD} (less than 0.6\% accuracy drop). At this $\Delta_\text{TH}$ value, the latency is reduced by $2.4\times$ to 6.9\,ms and energy consumption by $3.4\times$ to 36.11\,nJ/Decision.

These measurements were taken with the $\Delta$\ac{RNN} operating at a clock frequency of 125\,kHz. This highlights the potential for significant energy savings in \ac{KWS} applications through the optimization of $\Delta_\text{TH}$.
The measured \ac{SRAM} waveform (Fig.~\ref{fig:measure_3}) shows that the skew-resistant pre-charging scheme ensures $Q$ data is always updated near the falling clock edge.

\subsection{Comparison With SoTA Works}

Table~\ref{tab:Digital_FEx_comparison} compares the \ac{FEx} of our \ac{KWS} chip with other digital \ac{FEx} implementations \cite{Shan2020ISSCC, Giraldo2020JSSC}.
Our 65\,nm 12-bit digital serial \ac{IIR} \ac{BPF}-based \ac{FEx} occupies an area of 0.084\,mm$^2$, which is only 1.5$\times$ larger than the 28\,nm 8-bit digital serial \ac{FFT}-based \ac{MFCC} \ac{FEx} (0.057\,mm²) \cite{Shan2020ISSCC}. Despite the slightly larger area, our \ac{FEx} provides 2$\times$ larger feature dimensions (channels) and consumes only 1.22\,$\mu$W. 
It also achieves a 5.9$\times$ reduction in power consumption compared to a previous 65\,nm \ac{MFCC} \ac{FEx} \cite{Giraldo2020JSSC}, which also has lower feature precision (8-bit) compared to our 12-bit implementation.

Table~\ref{tab:KWS_comparison} compares our \ac{KWS} chip with other implementations ~\cite{Kim2022ISSCC,Seol2023,Frenkel2022ISSCC,Kosuge2023VLSI,tan202417}.
Kim et al.~\cite{Kim2022ISSCC} proposed a ring-oscillator-based time-domain \ac{FEx} for better technology scalability, however, their \ac{KWS} consumed 23\,$\mu$W power and achieved a limited 86.03\% accuracy for 10-keyword \ac{KWS}. 
Kosuge et al.~\cite{Kosuge2023VLSI} proposed a single-chip, fully synthesizable wired-logic \ac{DNN} processor that achieved 88\% accuracy for 10-keyword \ac{KWS} with a latency of only 1.2\,ms. However, the area of their chip is nearly 10$\times$ larger than our chip.
Tan et al.~\cite{tan202417} presented a low-power KWS chip also with a low decision latency of just 2\,ms, featuring a scalable 5T-SRAM.
Frenkel et al.~\cite{Frenkel2022ISSCC} proposed a spiking recurrent neural network processor that offers lower latency, however, it demonstrated results from only a single-keyword \ac{KWS}.
Seol et al.~\cite{Seol2023} presented a 1.5$\mu$W end-to-end \ac{KWS} \ac{SoC}, for only a 5-keyword \ac{KWS}.
These two works~\cite{Frenkel2022ISSCC, Seol2023} were limited to simpler tasks involving the classification of only a few keywords.


\begin{table*}[t]
    \centering
    \caption{Comparison of Digital FEx. implementations}
    \label{tab:Digital_FEx_comparison}
    \renewcommand{\arraystretch}{1.2}
    \begin{threeparttable}
        \begin{tabular}{|l|c|c|c|>{\columncolor{cyan!10}}c|}
            \hline
            \rowcolor{cyan!20}
            \textbf{Digital FEx.} & \textbf{Shan ISSCC'20 \cite{Shan2020ISSCC}} & \textbf{Giraldo JSSC'20 \cite{Giraldo2020JSSC}} & \textbf{Shan JSSC'23 \cite{Shan2022aadkws}}& \textbf{This Work} \\
            \hline
            Process (nm) & 28 & 65 & 28 & \textcolor{blue}{65} \\
            \hline
            Area (mm\(^2\)) & 0.057 & 0.66 & 0.093& \textcolor{blue}{0.084} \\
            \hline
            Clock (Hz) & 40k & 250k & 8k & \textcolor{blue}{128k} \\
            \hline
            Input Precision & 16 bits & 10 bits\(^A\) & 16 bits & \textcolor{blue}{12 bits} \\
            \hline
            Feature Precision & 8 bits & 8 bits  & 8 bits & \textcolor{blue}{12 bits} \\
            \hline
            Feature Type & MFCC & MFCC& MFCC & \textcolor{blue}{IIR} \\
            \hline
            Feature Dimension & 8 & $\leq$ 32 & 11& \textcolor{blue}{$\leq$ 16} \\
            \hline
            Backbone & 256-point FFT & 512-point FFT  & 128-point FFT & \textcolor{blue}{IIR-BPF} \\
            \hline
            Data Storage (bytes) & 256\(^D\)  &  - & 512\(^D\) & 200 \\
            \hline
            Freq. Range (Hz) & 16-8k & $\leq$ 8k & $\leq$ 4k& \textcolor{blue}{100-7.9k} \\
            \hline
            Supply (V) & 0.41 & 0.6/0.8  & 0.4 & \textcolor{blue}{0.65} \\
            \hline
            Power (\(\mu\)W) & 0.34 & 7.2\(^B\)  & 0.17& \textcolor{blue}{1.22\(^C\)} \\
            \hline
            Frame Shift (ms) & 16 & 16  & 32 & \textcolor{blue}{16} \\
            \hline
            Frame Window Length (ms) &32 & 32  & 32 & 16 \\
            \hline
            Serial FEx. & Yes & No& Yes & \textcolor{blue}{Yes} \\
            \hline
        \end{tabular}
        \begin{tablenotes}
            \footnotesize
            \item[\(^A\)] On-chip ADC
            \item[\(^B\)] Measured with 13-D features
            \item[\(^C\)] Measured with 10-D features
            \item[\(^D\)] Data storage for just FFT module
        \end{tablenotes}
    \end{threeparttable}
\end{table*}

\begin{table*}[t]
    \centering
    \caption{Comparison of KWS implementations}
    \label{tab:KWS_comparison}
    \renewcommand{\arraystretch}{1.2}
    \begin{threeparttable}
        \begin{tabular}{|l|c|c|c|c|c|c|c|c|}
            \hline
            \rowcolor{cyan!20}
            KWS  & \textbf{Kim } & \textbf{Frenkel } & \textbf{Seol } & \multicolumn{2}{c|}{\textbf{Kosuge}} & \textbf{Tan}& \multicolumn{2}{c|}{\textbf{This }} \\
            \rowcolor{cyan!20}
            Refs& \textbf{ISSCC'22 \cite{Kim2022ISSCC}} & \textbf{ISSCC'22~\cite{Frenkel2022ISSCC}} & \textbf{ISSCC'23 \cite{Seol2023}} & \multicolumn{2}{c|}{\textbf{VLSI'23 \cite{Kosuge2023VLSI}}} & \textbf{ISSCC'24 \cite{tan202417}}& \multicolumn{2}{c|}{\textbf{Work}} \\
            \hline
            
            Process (nm) & 65 & 28 & 28 & \multicolumn{2}{c|}{40} &28  & \multicolumn{2}{c|}{\textcolor{blue}{65}} \\
            \hline
            Supply (V) & 0.5/0.75 & 0.5 & 1.4/0.65/0.5 & \multicolumn{2}{c|}{0.5} & 0.35/0.9 & \multicolumn{2}{c|}{\textcolor{blue}{0.6/0.65}} \\
            \hline
            Area (mm\(^2\)) & 2.03 & 0.45 & 0.8 & \multicolumn{2}{c|}{7.63}& 0.121 &\multicolumn{2}{c|}{\textcolor{blue}{0.78}} \\
            \hline
            On-Chip Memory (kB) & 27 & 138 & 18 & \multicolumn{2}{c|}{0}& 16 &\multicolumn{2}{c|}{\textcolor{blue}{26.3}} \\
            \hline
            Clock (Hz) & 250k & 13M & 1M & \multicolumn{2}{c|}{120k}& 1M&\multicolumn{2}{c|}{\textcolor{blue}{125k}} \\
            \hline
             Feature Ex. & Analog Time & - & Digital Freq. &  \multicolumn{2}{c|}{TCN} & TCN & \multicolumn{2}{c|}{\textcolor{blue}{Digital Time}} \\
            \hline
           \multirow{2}{*}{Algorithm} & \multirow{2}{*}{RNN} & \multirow{2}{*}{Spiking RNN} & \multirow{2}{*}{Skip RNN} & \multicolumn{2}{c|}{\multirow{2}{*}{CNN}}& Transfer  & \multicolumn{2}{c|}{\textcolor{blue}{$\Delta$GRU-FC}} \\
           \cline{8-9} 
            & & & & \multicolumn{2}{c|}{}&Computing & \textcolor{blue}{$\Delta_\text{TH}=0$} & \textcolor{blue}{$\Delta_\text{TH}=0.2$} \\
            \hline
            Energy/Decision (nJ) & 285.20 & 42 & 23.68 & \multicolumn{2}{c|}{183.4}& 1.73& \textcolor{blue}{121.2} & \textcolor{blue}{36.11} \\
            \hline
            Comp. Latency (ms) & 12.4 & 5.7 & 16 & \multicolumn{2}{c|}{1.2}& 2 &\textcolor{blue}{16.4} & \textcolor{blue}{6.9} \\
            \hline
            KWS Power (\(\mu\)W) & 23\(^A\) & 79 & 1.48 \(^A\) & \multicolumn{2}{c|}{152.8} & 1.73 & \textcolor{blue}{7.36} & \textcolor{blue}{5.22} \\
            \hline
            Dataset & GSCD & SHD \(^B\) & GSCD & \multicolumn{2}{c|}{GSCD}&  GSCD & \multicolumn{2}{c|}{\textcolor{blue}{GSCD}} \\
            \hline
            \# Classes (\# Keywords) & 12 (10) & 2 (1) & 7 (5) & 35 (35) & 10 (10) & 12 (10) &\multicolumn{2}{c|}{\textcolor{blue}{11 (10) / 12 (10)}} \\
            \hline
            Accuracy (\%) & 86.03 & 90.7 & 92.8 & 78.2 &88.0 & 91.8 & \textcolor{blue}{91.1 / 90.1} & \textcolor{blue}{90.5 / 89.5} \\
            \hline
        \end{tabular}
        \begin{tablenotes}
            \footnotesize
            \item[\(^A\)] Power includes AFE, FE, and NN.
            \item[\(^B\)] Spiking Heidelberg Digits.
        \end{tablenotes}
    \end{threeparttable}
\end{table*}

\section{Conclusion}
\label{sec:conclusion}
This paper presented a \ac{KWS} chip featuring a fine-grained temporary sparsity-aware $\Delta$\ac{RNN} accelerator, a compact \ac{IIR} \ac{BPF}-based \ac{FEx}, and a low-power near-$V_\text{TH}$ \ac{SRAM}. With such techniques, our \ac{KWS} chip, fabricated in a 65\,nm CMOS process, achieves an energy consumption of 36\,nJ/decision with a classification accuracy of 90.5\%/89.5\% on the 11-class/12-class \ac{GSCD} dataset, respectively.
The bio-inspired $\Delta$\ac{RNN} accelerator leveraging temporal similarities between neighboring feature vectors extracted from input frames and network hidden states, eliminating unnecessary operations and memory accesses, achieving 2.4$\times$ latency reduction and 3.4$\times$ energy per decision reduction.
The serial \ac{IIR}-\ac{BPF} \ac{FEx} that features mixed-precision quantization, low-cost computing structure, and channel selection; achieving 5.7$\times$ power reduction and 4.7$\times$ area reduction.
The 24\,kB 0.6\,V near-threshold-operating full-custom \ac{SRAM} achieves 6.6$\times$ lower power consumption compared to the foundry-provided \ac{SRAM}.
The proposed \ac{KWS} chip, incorporating an \ac{IIR} \ac{BPF}-based \ac{FEx}, $\Delta$\ac{RNN} accelerator, and 24\,kB near-threshold \ac{SRAM}, occupies 0.78\,mm$^2$.


\section*{Acknowledgment}
The authors would like to thank Dr. Jinsu Lee, who was with KAIST, for the valuable feedback on the full-custom \ac{SRAM} design, and Harim Kim, who is with Samsung Electronics, for his technical support of running SPICE-based post-layout simulations in Cadence Virtuoso environment.

The authors acknowledge the Swiss National Science Foundation BRIDGE - Proof of Concept Project (40B1-0 213731) and CA-DNNEdge project (208227) for partial funding of this work.

\bibliographystyle{IEEEtran}
\bibliography{IEEEabrv,ref}

\begin{IEEEbiography}[{\includegraphics[width=1in,height=1.25in,clip,keepaspectratio]{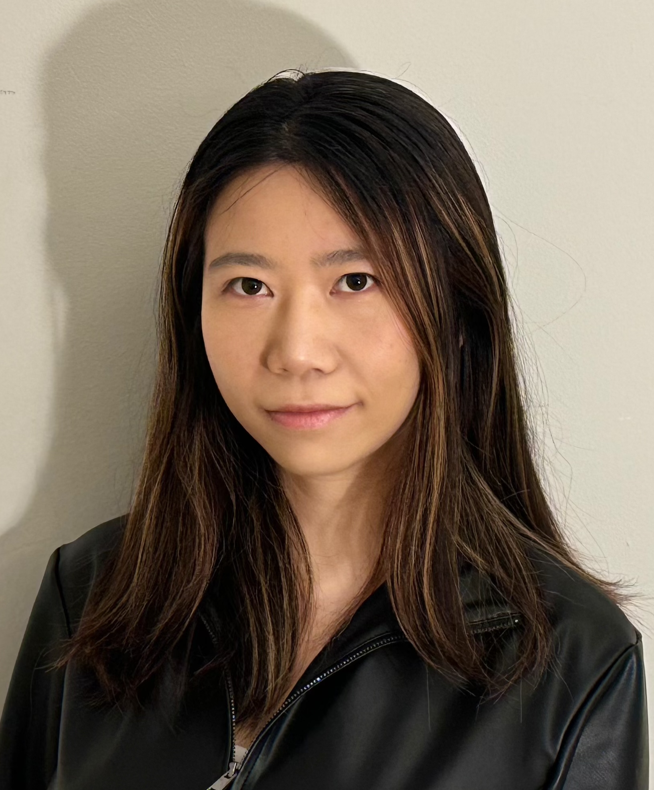}}]{Qinyu Chen} (Member, IEEE) is an Assistant Professor at the Leiden Institute of Advanced Computer Science (LIACS), Leiden University, Leiden, the Netherlands. She received the Ph.D. degree in Electronic Science and Technology from Nanjing University, Nanjing, China, in 2021. She was a visiting student from 2019 to 2020, and a Postdoctoral Researcher from 2022 to 2024, at the Institute of Neuroinformatics, University of Zürich and ETH Zürich, Zurich, Switzerland. 
Her current research interest includes the seamless brain-inspired AI system at the edge, and its application in healthcare, AR/VR with a focus on event-based processing. In 2022, She received a Bridge Fellowship Grant from the Swiss National Science Foundation (SNSF) and Innosuisse. She also serves as a member of the Neural Systems and Applications (NSA) Technical Committee in the IEEE Circuit and System Society (CASS). 
\end{IEEEbiography}

\begin{IEEEbiography}[{\includegraphics[width=1in,height=1.25in,clip,keepaspectratio]{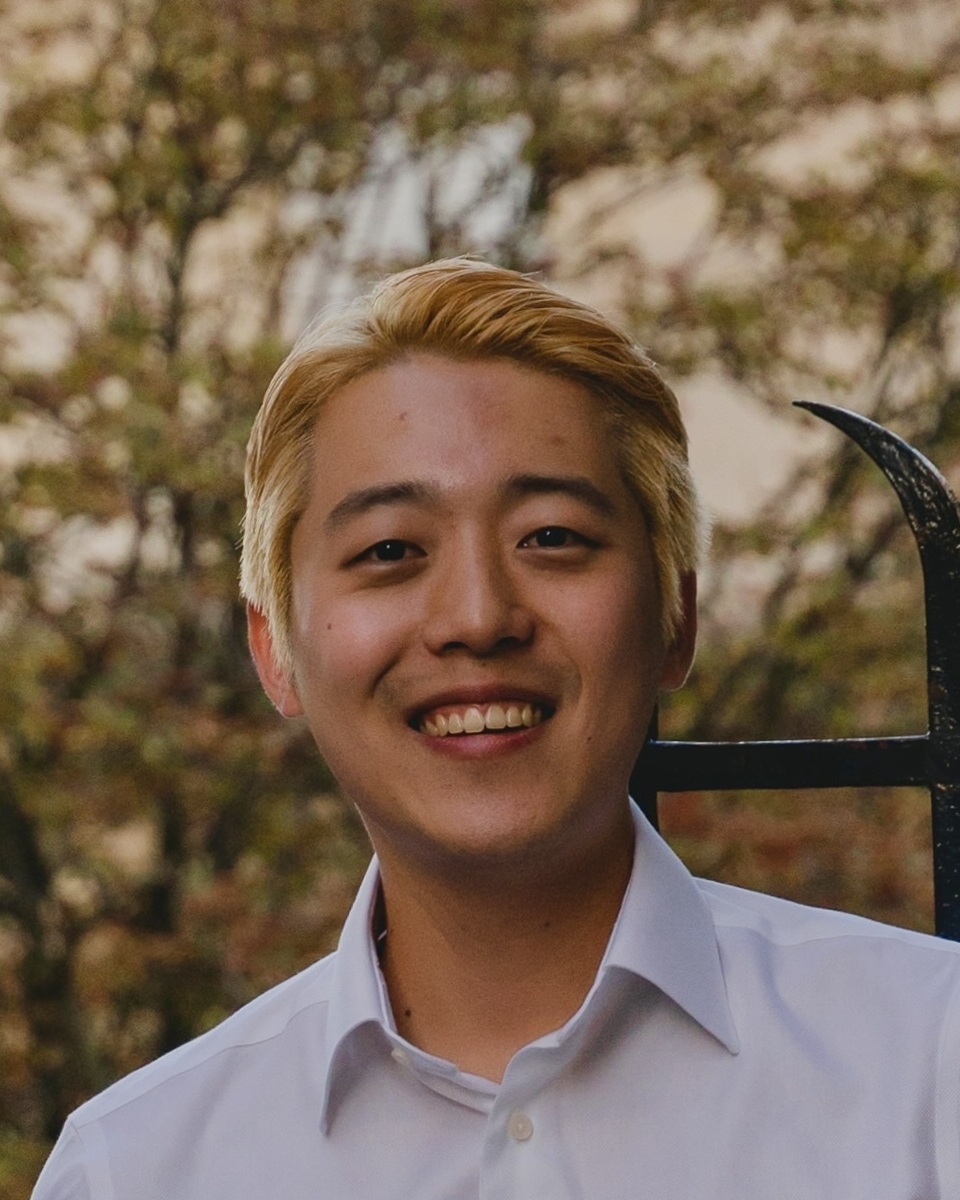}}]{Kwantae Kim} (Member, IEEE) received the B.S., M.S., and Ph.D. degrees in the School of Electrical Engineering from KAIST, South Korea, in 2015, 2017, and 2021, respectively. He is an Assistant Professor at the Department of Electronics and Nanoengineering, School of Electrical Engineering, Aalto University, Finland.

From 2015 to 2017, he was also with Healthrian R\&D Center, South Korea, where he designed analog readout ICs for mobile healthcare solutions. He was a Visiting Student in 2020, and a Postdoctoral Researcher from 2021 to 2023, at the Institute of Neuroinformatics, University of Zurich and ETH Zurich, and an Established Researcher from 2023 to 2024, at the Department of Information Technology and Electrical Engineering (D-ITET), ETH Zurich, Switzerland. His research interests include analog/mixed-signal ICs and full-custom memory ICs for neuromorphic signal processing, biomedical sensors, and in-memory computing.
\end{IEEEbiography}

\begin{IEEEbiography}[{\includegraphics[width=1in,height=1.25in,clip,keepaspectratio]{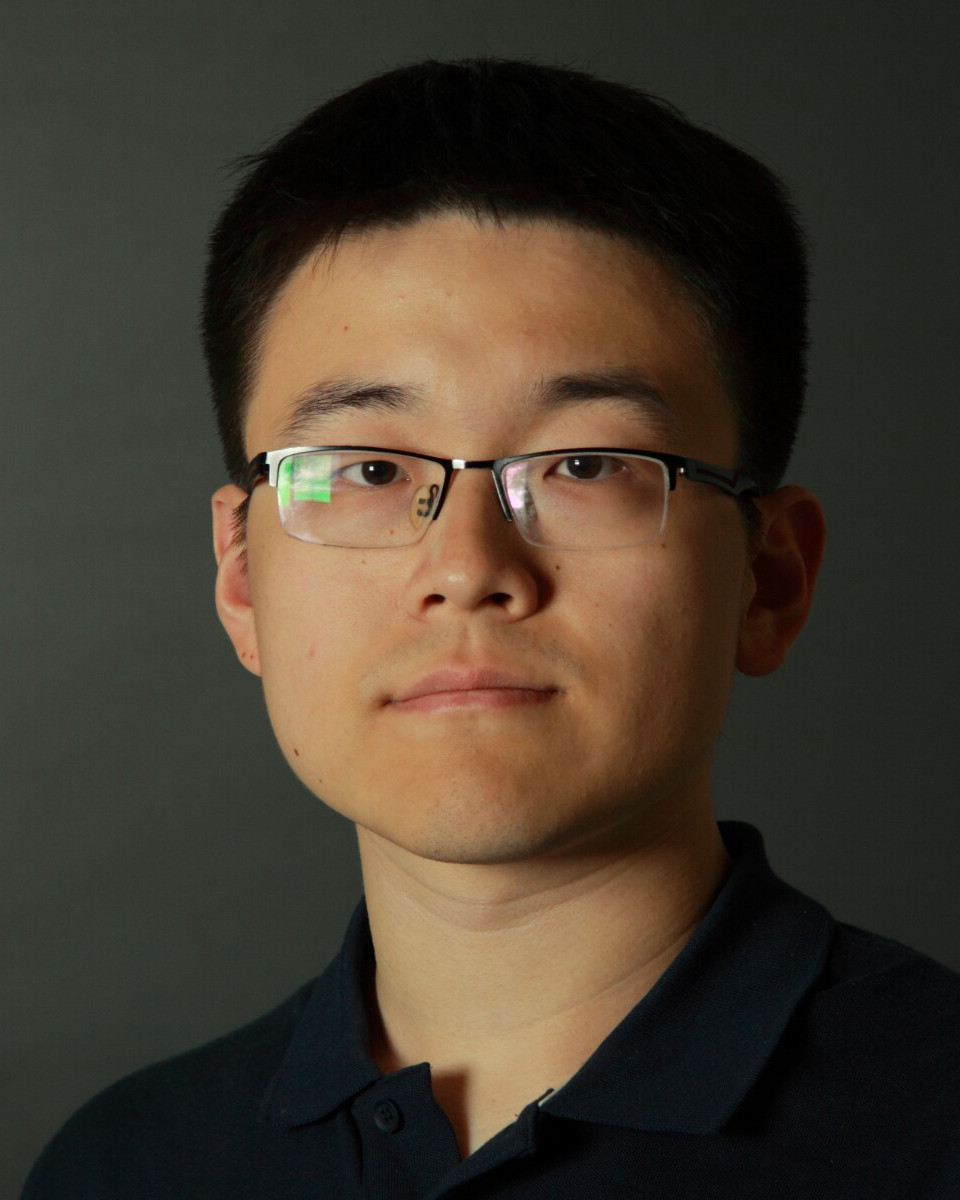}}]{Chang Gao} (Member, IEEE) received his Ph.D. degree with Distinction in Neuroscience from the Institute of Neuroinformatics, University of Zürich and ETH Zürich, Zürich, Switzerland, in March 2022. In August 2022, he joined the Delft University of Technology, The Netherlands, as an Assistant Professor in the Department of Microelectronics. He received the 2022 Misha Mahowald Early Career Award in Neuromorphic Engineering, the 2022 Marie-Curie Postdoctoral Fellowship, and the title of 2023 MIT Technology Review Innovators Under 35 in Europe for his contribution to algorithm-hardware co-design for efficient sparse recurrent neural network edge computing. His current research interest is in developing digital AI hardware accelerators and their applications in future wireless communications, video/audio processing, healthcare, and robotics.
\end{IEEEbiography}

\begin{IEEEbiography}[{\includegraphics[width=1in,height=1.25in,clip,keepaspectratio]{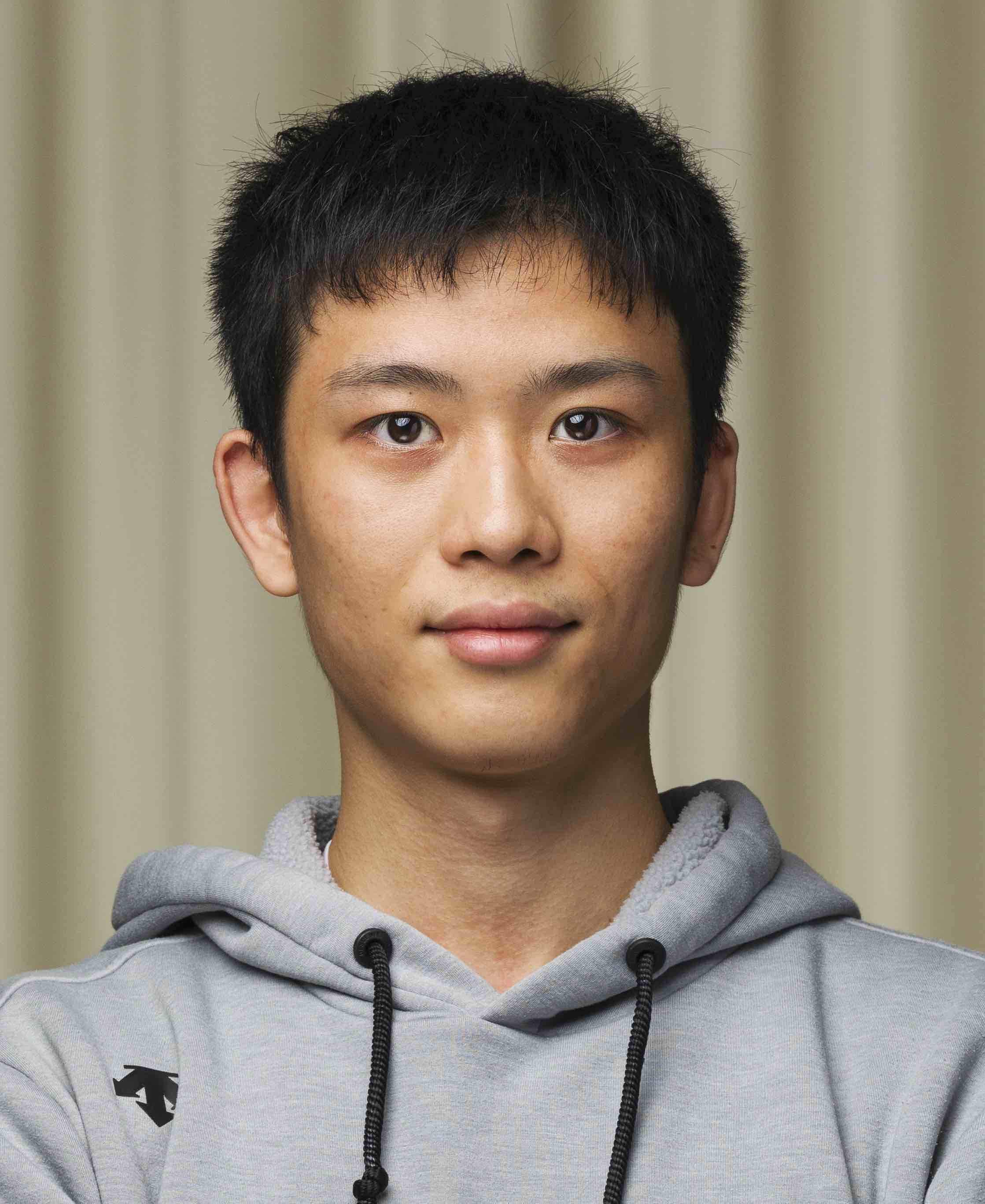}}]{Sheng Zhou} (Graduate Student Member, IEEE) received his bachelor's degree in computer science from the Hong Kong University of Science and Technology, and master's degree in data science from ETH Zurich. He is currently pursuing the Ph.D. degree in the Sensors Group at the Institute of Neuroinformatics, University of Zurich and ETH Zurich. His research interest is in mixed-signal circuit design for ultra-low-power edge applications.
\end{IEEEbiography}

\begin{IEEEbiography}[{\includegraphics[width=1in,height=1.25in,clip,keepaspectratio]{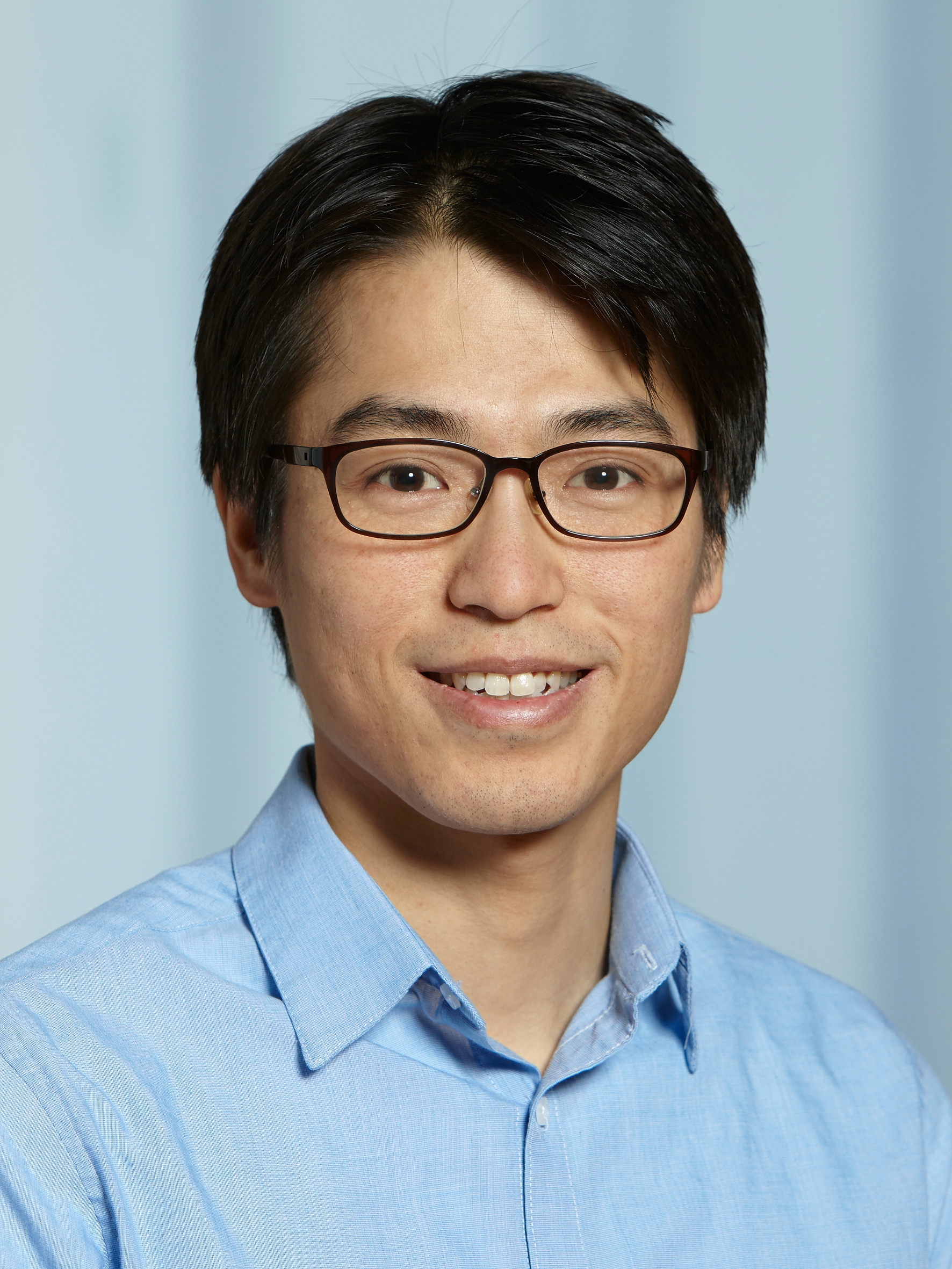}}]{Taekwang Jang} (Senior Member, IEEE) received his B.S. and M.S. in electrical engineering from KAIST, Korea, in 2006 and 2008, respectively. From 2008 to 2013, he worked at Samsung Electronics Company Ltd., Yongin, Korea, focusing on mixed-signal circuit design, including analog and all-digital phase-locked loops for communication systems and mobile processors. In 2017, he received his Ph.D. from the University of Michigan and worked as a post-doctoral research fellow at the same institution. In 2018, he joined ETH Zürich as an assistant professor and is leading the Energy-Efficient Circuits and Intelligent Systems group. He is also a member of the Competence Center for Rehabilitation Engineering and Science, and the chair of the IEEE Solid-State Circuits Society, Switzerland chapter.

His research focuses on circuits and systems for highly energy-constrained applications such as wireless sensor nodes and biomedical interfaces. Essential building blocks such as a sensor interface, energy harvester, power converter, communication transceiver, frequency synthesizer, and data converters are his primary interests. He holds 15 patents and has (co)authored more than 80 peer-reviewed conferences and journal articles. He is the recipient of the 2024 IEEE Solid-State Circuits Society New Frontier Award, the SNSF Starting Grant, the IEEE ISSCC 2021 and 2022 Jan Van Vessem Award for Outstanding European Paper, the IEEE ISSCC 2022 Outstanding Forum Speaker Award, and the 2009 IEEE CAS Guillemin-Cauer Best Paper Award. Since 2022, he has been a TPC member of the IEEE International Solid-State Circuits Conference (ISSCC), IMD Subcommittee, and IEEE Asian Solid-State Circuits Conference (ASSCC), Analog Subcommittee. He also chaired the 2022 IEEE International Symposium on Radio-Frequency Integration Technology (RFIT), Frequency Generation Subcommittee. Since 2023, he has been serving as an Associate Editor for the Journal of Solid-State Circuits (JSSC) and was appointed as a Distinguished Lecturer for the Solid-State Circuits Society in 2024.

\end{IEEEbiography}

\begin{IEEEbiography}[{\includegraphics[width=1in,height=1.25in,clip,keepaspectratio]{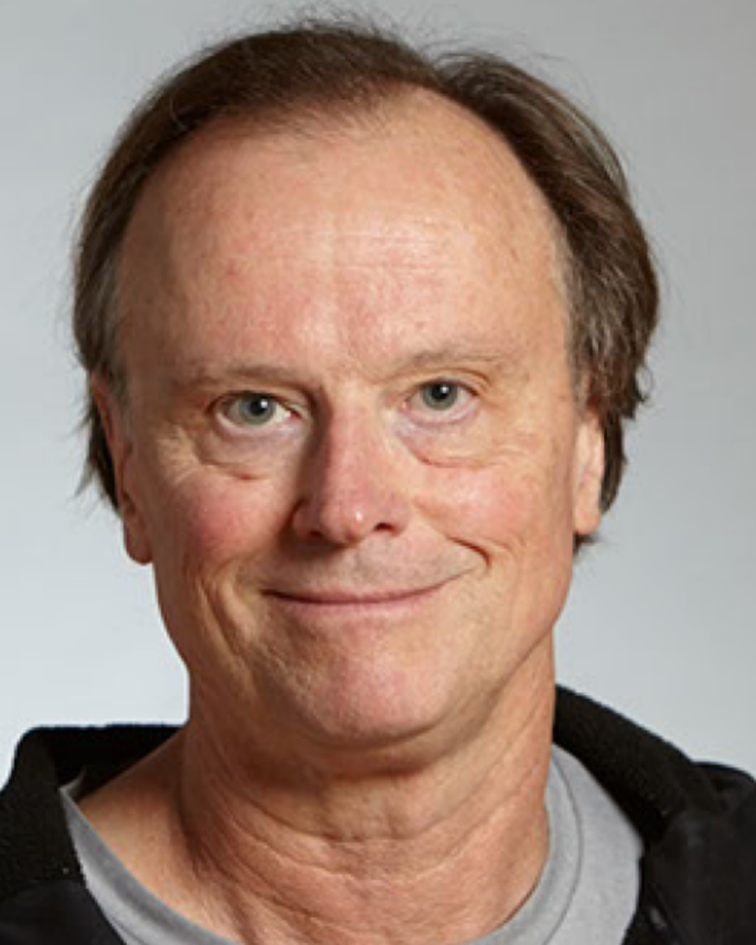}}]{Tobi Delbruck} (Fellow, IEEE) 
received the degree in physics from University of California in 1986 and Ph.D. degree from Caltech in 1993. Currently, he is a Professor of Physics and Electrical Engineering at the Institute of Neuroinformatics, University of Zurich and ETH Zurich, where he has been since 1998. He co-directs the Sensors group, and his current research focus is on neuromorphic sensory processing, control, and efficient hardware AI.
\end{IEEEbiography}

\begin{IEEEbiography}[{\includegraphics[width=1in,height=1.25in,clip,keepaspectratio]{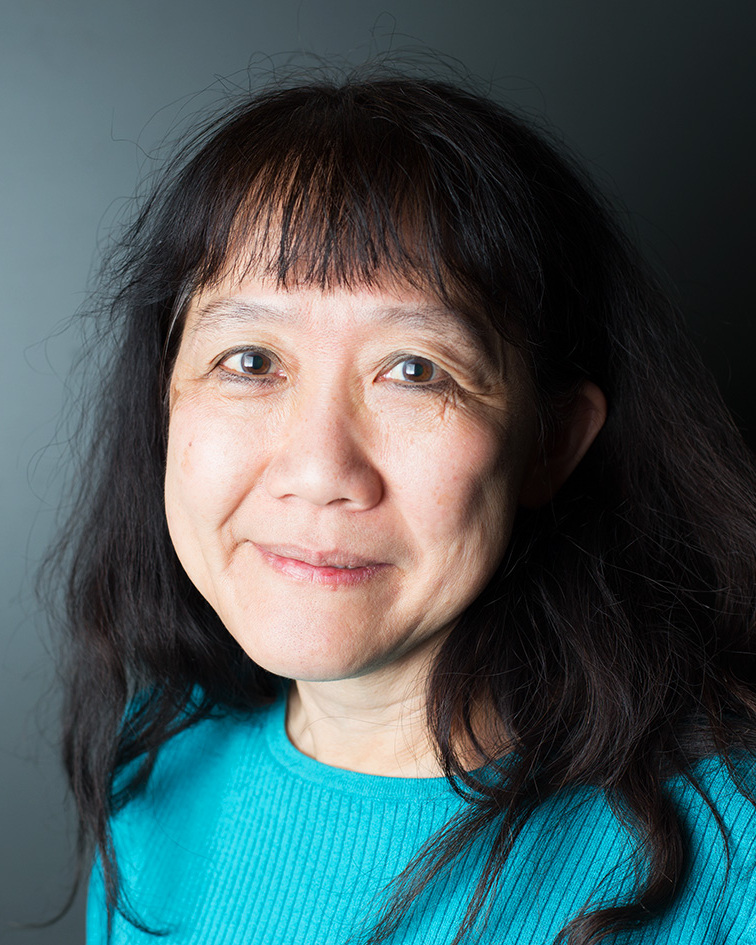}}]{Shih-Chii Liu} (Fellow, IEEE) 
received the bachelor’s degree in electrical engineering from the Massachusetts Institute of Technology, Cambridge, MA, USA, and the Ph.D. degree in the Computation and Neural Systems program from the California Institute of Technology, Pasadena, CA, USA, in 1997. She is currently Adjunct Professor in the Faculty of Science at the University of Zurich. She co-directs the Sensors group at the Institute of Neuroinformatics, University of Zurich and ETH Zurich. Her group’s research focuses on sensor integrated circuit designs including the spiking silicon cochlea and bio-inspired auditory sensors; and real-time energy-efficient hardware systems that combine both sensor and event-driven low-compute deep neural network algorithms, targeting always-on edge AI and wearable applications.

\end{IEEEbiography}

\vfill

\end{document}